\documentclass[aps,pra,twocolumn,reprint,groupedaddress,showpacs]{revtex4-1}

\usepackage{amsmath}
\usepackage{amssymb}
\usepackage{graphicx}
\usepackage{color} 
\usepackage[normalem]{ulem} 
\DeclareMathAlphabet{\mathpzc}{OT1}{pzc}{m}{it}

\newcommand{\ket}[1]{\left|#1\right\rangle}

\begin{document}


\title{Optimal post-processing for a generic single-shot qubit readout}


\author{B.~D'Anjou}
\author{W.A.~Coish}
\affiliation{Department of Physics, McGill University, Montreal, Quebec H3A 2T8, Canada}


\date{\today}

\begin{abstract}
We analyze three different post-processing methods applied to a single-shot qubit readout: the average-signal (boxcar filter), peak-signal, and maximum-likelihood methods. In contrast to previous work, we account for a stochastic turn-on time $t_i$ associated with the leading edge of a pulse signaling one of the qubit states. This model is relevant to spin-qubit readouts based on spin-to-charge conversion and would be generically reached in the limit of large signal-to-noise ratio $r$ for several other physical systems, including fluorescence-based readouts of ion-trap qubits and nitrogen-vacancy center spins. We derive analytical closed-form expressions for the conditional probability distributions associated with the peak-signal and boxcar filters. For the boxcar filter, we find an asymptotic scaling of the single-shot error rate $\varepsilon \sim \ln r/\sqrt{r}$ when $t_i$ is stochastic, in contrast to the result $\varepsilon \sim \ln r/ r$ for deterministic $t_i$. Consequently, the peak-signal method outperforms the boxcar filter significantly when $t_i$ is stochastic, but is only marginally better for deterministic $t_i$ (a result that is consistent with the widespread use of the boxcar filter for fluorescence-based readouts and the peak-signal for spin-to-charge conversion). We generalize the theoretically optimal maximum-likelihood method to stochastic $t_i$ and show numerically that a stochastic turn-on time $t_i$ will always result in a larger single-shot error rate. Based on this observation, we propose a general strategy to improve the quality of single-shot readouts by forcing $t_i$ to be deterministic. 
\end{abstract}


\maketitle

\section{Introduction \label{sec:Introduction}}

\begin{figure}
\centering
\includegraphics[width=\columnwidth]{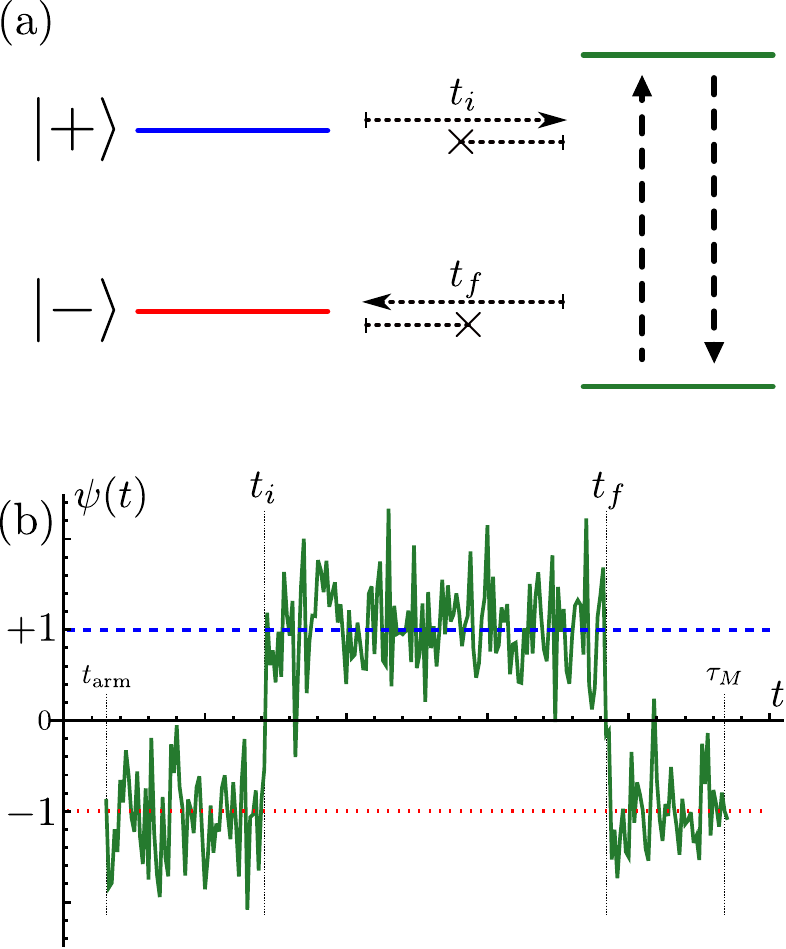}
\centering
\caption{(Color online) {\bf(a)} Schematic representation of a generic cycling process. This diagram represents, e.g., the fluorescence cycle of an NV-center or trapped ion, or the flow of electrons through an SET or QPC during spin-to-charge conversion. If the system is initially in the excited state $\ket{+}$ (dashed blue line) at $t=0$, it can trigger cycling between two cycling states (solid green line) at time $t_i$. The cycling ends at time $t_f$ when the system falls into the ground state $\ket{-}$ (dotted red line). If the system is initially in the state $\ket{-}$, cycling cannot occur because of either selection rules or energy conservation requirements, as indicated by the crosses. {\bf(b)} Noisy time-dependent signal $\psi(t)$ resulting from the cycling process of (a) when the initial state is $\ket{+}$. The readout phase starts at $t=0$ and acquisition starts after an arming time $t_{\mathrm{arm}}$. Initially, cycling does not occur and the signal takes the average value $\left\langle\psi\right\rangle = -1$. At a random time $t_i$, cycling is triggered and the signal rises to $\left\langle\psi\right\rangle = 1$. At a subsequent random time $t_f$, cycling stops and the signal again drops to $\left\langle\psi\right\rangle = -1$. Acquisition stops after a measurement time $\tau_M$. Throughout, we assume that $t_{\mathrm{arm}}\rightarrow 0$, that $t_i$ and $\tau=t_f-t_i$ follow Poisson processes and that the noise is white and Gaussian. \label{fig:fig1}}
\end{figure}

A prerequisite for many quantum information processing applications is the ability to perform a strong projective single-shot measurement of a quantum bit (qubit) in the computational basis, $\left\{\ket{+},\ket{-}\right\}$~\cite{sohn1997}. In general, the readout procedure depends on the particular measurement apparatus and physical system used to encode the qubit. However, a wide variety of high-fidelity single-shot readouts rely on the conditional amplification of one of the qubit states, say $\ket{+}$, by means of a cycling process [see Fig.~\ref{fig:fig1}(a)]. For example, the state of a nitrogen-vacancy center (NV-center) in diamond~\cite{neumann2010,robledo2011,dreau2013}, of trapped ions~\cite{myerson2008}, and of quantum-dot spins~\cite{vamivakas2010,delteil2013,puri2013} can be mapped to a fluorescence signal when the system is driven by a laser. Similarly, the spin state of electrons in semiconductor quantum dots or phosphorus donors in silicon can be mapped to a current through a nearby quantum point contact (QPC) or single-electron transistor (SET) via spin-to-charge conversion~\cite{engel2004,elzerman2004,morello2010,nowack2011,simmons2011,pla2013}. Hybrid optical/electrical
approaches to single-spin readout have also been demonstrated~\cite{yin2013}. Readouts of semiconductor singlet-triplet qubits~\cite{hanson2005,johnson2005,petta2006,barthel2009,prance2012} and superconducting qubits~\cite{lin2013} also rely on similar amplification mechanisms. These cycling processes result in a time-dependent analog signal $\psi(t)$ related to the number of cycles per unit time. If cycling is observed, it is inferred that the qubit state must have been $\ket{+}$; otherwise, it is inferred that the qubit state must have been $\ket{-}$. Such readouts are quantum non-demolition (QND) since each cycle preserves the information associated with the initial qubit state. The fact that these readouts are QND is one reason they can reach high fidelities. To convert analog information associated with the noisy cycling signal $\psi(t)$, a post-processing procedure must be chosen to minimize the frequency of errors in the assignment of the binary state. Although higher-level and hardware-independent protocols can be used to minimize the uncertainty in quantum process tomography~\cite{medford2013,merkel2013}, an understanding of the dependence of the error rate on the underlying physical parameters is useful in further improving state reconstruction.

Interestingly, different communities have used different post-processing protocols to optimize the fidelity of their readouts, even though the various physical readouts are based on very similar cycling processes. In particular, fluorescence-based experiments have typically relied on integrating the signal $\psi(t)$ over time to detect cycling (the so-called \emph{boxcar} filter)~\cite{myerson2008,neumann2010,robledo2011,dreau2013}, although more sophisticated Bayesian inference procedures have also been used~\cite{myerson2008,sun2013}. Spin-to-charge conversion experiments for semiconductor spin qubits, on the other hand, have typically been analyzed through a measurement of the peak of the signal $\psi(t)$ (the \emph{peak-signal} filter)~\cite{elzerman2004,morello2010,simmons2011,pla2013}. In light of the striking similarities between these readouts, it is natural to ask whether the disparity in post-processing originates from a fundamental difference between the experiments. In fact, the only qualitative difference between the two cases is the mechanism triggering the cycling process. For spin-to-charge conversion, there is a random turn-on time $t_i$ after the beginning of the readout phase, where $t_i$ follows a Poisson process. In contrast, for fluorescence-based readouts cycling typically starts on a very short time scale $t_i\approx 0$ as driving is turned on. Here we demonstrate that the uncertainty resulting from this stochastic turn-on time indeed accounts for the disparity in post-processing procedures, and we propose an avenue for increasing the fidelity of such readouts by making the turn-on time deterministic.

Quantum measurements based on cycling processes have been the subject of various theoretical studies. In particular, considerable attention has been given to the readout of a qubit using a non-QND cycling process (with back-action on the qubit) for several distinct physical systems~\cite{gurvitz1997,shnirman1998,makhlin2000,korotkov2001,gurvitz2005,gilad2006,jiao2007,kreisbeck2010}. Protocols for the optimal readout of a qubit using a QND cycling process have been studied in great detail in Ref.~\cite{gambetta2007} for the case $t_i=0$, although the statistics for the peak signal were not derived in that work. In this paper we analytically obtain the statistics of the peak signal in the case of a stochastic turn-on time $t_i$ and Gaussian white noise. We demonstrate the validity of our approach by fitting our analytical probability distribution for the peak signal to that measured in the readout of single spin qubits in silicon in Ref.~\cite{morello2010}. Most importantly, we show that a significant improvement in fidelity is obtained by using the peak-signal filter over the boxcar filter if the turn-on time $t_i$ is stochastic. More precisely, we prove that for large signal-to-noise ratio $r$, the boxcar-filter error rate $\varepsilon$ scales like $\varepsilon \sim \ln r/r$ when $t_i$ is fixed while the error rate scales like $\varepsilon \sim \ln r /\sqrt{r}$ when $t_i$ follows a Poisson process. The key observation is that the loss of information associated with a stochastic $t_i$ can be largely compensated by using a simple peak-signal filter instead of a boxcar filter. This result explains the disparity between post-processing methods used in different experiments and indicates which method should be used in future experiments. Furthermore, we generalize the optimal maximum-likelihood filter developed in Ref.~\cite{gambetta2007} and show numerically that the stochasticity of $t_i$ reduces the fidelity significantly even when this theoretically optimal procedure is followed. This result leads to the conclusion that physical readouts with stochastic turn-on time $t_i$ can be generically improved by engineering them such that $t_i\simeq 0$ becomes deterministic.

The remainder of the text is organized as follows. In Sec.~\ref{sec:errorRate}, we express the readout error rate in terms of the probability distributions of measured observables. In Sec.~\ref{sec:modelSignal}, we introduce a model for the noisy cycling signal and formally define the peak-signal and boxcar filters. In Sec.~\ref{sec:statistics}, we analytically derive the conditional probability distributions for the observables derived from the peak-signal and boxcar filters. We then fit the peak-signal distributions to experimental data presented in Ref.~\cite{morello2010}. In Sec.~\ref{sec:errorRateAnalysis}, we numerically obtain the error rate for the peak-signal and boxcar filters and analytically derive the scaling of the boxcar-filter error rate for large signal-to-noise ratio. Finally, in Sec.~\ref{sec:bayesian} we generalize the maximum-likelihood filter of Ref.~\cite{gambetta2007} to the case of stochastic $t_i$. We conclude in Sec.~\ref{sec:conclusions}.

\section{Error rate\label{sec:errorRate}}

For the most general readout, the goal is to infer the initial qubit state from some observable $O$. For example, $O$ could be the peak $\psi_p$ (obtained from the peak-signal filter) or the time average $\bar{\psi}$ (obtained from the boxcar filter) of some analog signal $\psi(t)$. These quantities are defined in Eqs.~\eqref{eq:peakSignal} and \eqref{eq:boxCar}, below. In Sec.~\ref{sec:bayesian}, we will take $O$ to be the full measurement record $\psi(t)$ (appropriate for the maximum-likelihood filter). To infer the state, we define the likelihood ratio~\cite{tsang2012}:
\begin{align}
	\Lambda = \frac{P(O|+)}{P(O|-)} = \frac{P(-)}{P(+)}\times \frac{P(+|O)}{P(-|O)}, \label{eq:likelihoodRatio}
\end{align}
where $P(O|\pm)$ is the probability density of measuring the observable $O$ given the state $\ket{\pm}$ and where the last equality is obtained using Bayes' theorem. If $\Lambda$ is greater than the threshold $\lambda = P(-)/P(+)$, the state is most likely $\ket{+}$; otherwise, the state is most likely $\ket{-}$. For simplicity, we assume that the prior probabilities for the initial state are balanced, $P(\pm)=1/2$, in which case the threshold is $\lambda = 1$. The average error rate is then given by:
\begin{align}
 \varepsilon = \frac{1}{2}(\varepsilon_+ + \varepsilon_-), \label{eq:averageErrorRate}
\end{align}
where	$\varepsilon_{+} = P(\Lambda < 1 | +)$ and	$\varepsilon_{-} = P(\Lambda > 1 | -)$ are the error rates conditional on the initial qubit state. These expressions are valid for an arbitrary observable $O$. 

In the common case where the observable $O$ is a real scalar, as is the case for the peak-signal and boxcar filters, the threshold $\lambda=1$ is equivalent to a threshold $\nu$ for $O$, satisfying $P(\nu|+)=P(\nu|-)$. The conditional error rates are then given by~\cite{gambetta2007,barthel2009,tsang2012}:
\begin{align}
	\varepsilon_+ = \int_{-\infty}^{\nu} dO\,P(O|+)\; , \;\; \varepsilon_- = \int_{\nu}^{\infty} dO\,P(O|-), \label{eq:conditionalErrorRates}
\end{align}
and the fidelity is simply $F=1-\varepsilon$. For the maximum-likelihood filter, such simple thresholding is not possible since $O$ is a multi-dimensional object, namely the signal $\psi(t)$ given at all times $t$. In this case, the error rate \eqref{eq:averageErrorRate} must be obtained from Monte-Carlo simulations (see Sec.~\ref{sec:bayesian})~\cite{gambetta2007}.

\section{Model of the signal and noise \label{sec:modelSignal}}

\subsection{Noisy cycling signal \label{sec:cyclingSignal}}

We now model the time-dependent signal $\psi(t)$ resulting from the cycling process of Fig.~\ref{fig:fig1}(a). This could be, for example, a cycling fluorescence transition in NV-centers and ion traps, or the current flowing through an SET or QPC in the case of spin-to-charge conversion in semiconductor spin qubits. If the qubit is initially in the ground state $\ket{-}$, cycling does not occur. Thus, the average of $\psi(t)$ over realizations of the noise is the same at all times $t$. We choose the convention that (ensemble averages are indicated by angular brackets throughout):
\begin{align}
	\left\langle \psi(t) \right\rangle = -1. \label{eq:averageSignalMinus}
\end{align}
If the qubit is initially in the excited state $\ket{+}$, cycling begins at a random turn-on time $t_i$ and ends at a random turn-off time $t_f$. The stochasticity of $t_i$ and $t_f$ typically results from coupling the qubit states $\ket{\pm}$ to a broadband continuum (the radiation field in the case of a fluorescence readout~\cite{myerson2008,neumann2010,vamivakas2010,robledo2011}, or a Fermi sea of electronic states in the case of spin-to-charge conversion~\cite{elzerman2004,morello2010,simmons2011}), leading to a Markovian process, hence a Poissonian (exponential) distribution of $t_i$ and $t_f$. As illustrated in Fig.~\ref{fig:fig1}(b), the result is a noisy time-dependent signal $\psi(t)$ such that:
\begin{align}
	\left\langle \psi(t) \right\rangle = 2\left[\theta\left(t-t_i\right)-\theta\left(t-t_f\right)\right]-1. \label{eq:averageSignalPlus}
\end{align}
Here, the turn-on time $t_i$ and pulse width $\tau=t_f-t_i$ each follows an independent Poisson process. Therefore, the probability distribution for $t_i$ and $t_f$ has the exponential form:
\begin{align}
	P(t_i,t_f) = \Gamma e^{-\Gamma t_i} e^{-(t_f-t_i)}. \label{eq:probabilitytitf}
\end{align}
Here and throughout, time is measured in units of the average pulse width $\left\langle \tau\right\rangle$ and $\Gamma$ is the ratio of $\left\langle \tau\right\rangle$ to the average turn-on time $\left\langle t_i\right\rangle$. We recover the case of a deterministic turn-on time $t_i\rightarrow 0$ when $\Gamma\rightarrow\infty$. This is typically the relevant case for fluorescence-based readouts~\cite{myerson2008,neumann2010,vamivakas2010,robledo2011}. As indicated in Fig.~\ref{fig:fig1}(b), the stochastic turn-on time $t_i$ must be distinguished from a deterministic arming time $t_{\mathrm{arm}}$~\cite{elzerman2004,gambetta2007} during which the qubit may relax [see Fig.~\ref{fig:fig1}(b)]. Indeed, the uncertainty in $t_i$ will affect the readout error rate \emph{even if} the qubit relaxation time is infinite or $t_{\mathrm{arm}}=0$. In the following analysis we will neglect qubit relaxation and take the arming time to be negligible, $t_\mathrm{arm}\simeq 0$.

For simplicity, we also assume that $\psi(t)$ is subject to Gaussian white noise, i.e. that the signal autocorrelation function is:
\begin{align}
	\left\langle \delta\psi(t)\delta\psi(t')\right\rangle = r^{-1} \delta(t-t'), \label{eq:noiseAutocorrelation}
\end{align}
where $\delta\psi(t)=\psi(t)-\left\langle\psi(t)\right\rangle$. Here, $r$ is the (power) signal-to-noise ratio integrated over an interval $\left\langle\tau\right\rangle=1$:
\begin{align}
	r^{-1}=\int_{0}^{1}\!\!\!\!dt\int_{0}^{1}\!\!\!\!dt' \left\langle \delta\psi(t)\delta\psi(t')\right\rangle.
\end{align}
The assumption of Gaussian noise is only valid when the number of cycling events is much larger than one, so that we can treat $\psi(t)$ as a continuous variable. Furthermore, for simplicity we assume shot noise is negligible compared to other sources of stationary Gaussian white noise (e.g. due to amplifier electronics). In the opposite limit where the readout is limited by the shot-noise power, the error rate is simply given by the probability that no cycling event occurs~\cite{robledo2011}. 

\subsection{Peak-signal and boxcar filters \label{sec:modelFilters}}

We take the signal $\psi(t)$ to be measured during a time $\tau_M$ [see Fig.~\ref{fig:fig1}(b)]. In practice, each data point on such a trace is necessarily acquired over a finite bin time $\tau_b$, corresponding to the inverse bandwidth of either a measurement apparatus or of a low-pass filter applied for post-processing. For simplicity, we assume that $\tau_M$ is separated in $N$ bins of length $\tau_b=\tau_M/N$ (see Fig.~\ref{fig:fig2}). The $l^{th}$ bin, starting at time $l \tau_b$, is then assigned its time-averaged value:
\begin{align}
\bar{\psi}_l = \frac{1}{\tau_b}\int_{l\tau_b}^{(l+1)\tau_b}\!\!\!\!dt\,\psi(t)\;,\;\;l=0,1,2,.,N-1. \label{eq:binSignal}
\end{align}
With Gaussian white noise, Eq.~\eqref{eq:noiseAutocorrelation}, the probability distribution $p_{\phi}(\bar{\psi}_l) \equiv P(\bar{\psi}_l|\pm;t_i,t_f)$ for $\bar{\psi}_l$ in bin $l$ is:
\begin{align}
	p_{\phi}(\bar{\psi}_l)= \mathcal{N}_{\sigma}\left( \bar{\psi}_l-\phi\right) = \frac{1}{\sqrt{2 \pi \sigma^2}} e^{-\frac{\left(\bar{\psi}_l-\phi\right)^2}{2\sigma^2}}, \label{eq:binDistribution}
\end{align}
where $\mathcal{N}_{\sigma}$ is the normal distribution of zero mean and of variance $\sigma^2=(r\tau_b)^{-1}$. Here, $\phi\equiv\left\langle\bar{\psi}_l\right\rangle$ is the average of $\bar{\psi}_l$ over realizations of the noise. It will also be useful to define the cumulative distribution function $q_{\phi}(\bar{\psi}_l)$ corresponding to $p_{\phi}$:
\begin{align}
	q_{\phi}(\bar{\psi}_l)=\int_{-\infty}^{\bar{\psi}_l}d\psi\;p_{\phi}(\psi) = \frac{1}{2}\mathrm{erfc}\left(\frac{ \phi - \bar{\psi}_l }{\sqrt{2\sigma^2}}\right). \label{eq:cumulativeBinDistribution}
\end{align}
We can now define the peak signal $\psi_p$ on $\tau_M$ as the maximum of $\bar{\psi}_l$ over all bins:
\begin{align}
	\psi_p = \max_{l<N}\,\bar{\psi}_l. \label{eq:peakSignal}
\end{align}
The time-averaged signal $\bar{\psi}$, corresponding to the boxcar filter, is then recovered as a special case of the peak signal with $N=1$:
\begin{align}
	\bar{\psi}=\frac{1}{\tau_M}\int_{0}^{\tau_M}\!\!\!\!dt\,\psi(t). \label{eq:boxCar}
\end{align}
Note that in the end, the error rate \eqref{eq:averageErrorRate} must be optimized with respect to both the bin time $\tau_b$ and the measurement time $\tau_M$, in addition to the threshold $\nu$.

The form \eqref{eq:averageSignalPlus} of the signal suggests an alternative two-time boxcar filter of the form $(\tau_{M2}-\tau_{M1})^{-1}\int_{\tau_{M1}}^{\tau_{M2}}\!\!dt\,\psi(t)$, where both $\tau_{M1}$ and $\tau_{M2}$ must be optimized. However, we have verified numerically that this two-time boxcar filter leads to a negligible improvement on the error rate of the simple boxcar filter, Eq.~\eqref{eq:boxCar}, for reasons that we detail in Sec.~\ref{sec:errorRateAnalysis}. Thus, in the following we only consider the simple boxcar filter defined in Eq.~\eqref{eq:boxCar}.

\section{Statistics of the peak-signal and boxcar filters \label{sec:statistics}}

To obtain the error rate for the peak-signal and boxcar filters, Eqs.~\eqref{eq:averageErrorRate} and \eqref{eq:conditionalErrorRates}, we must first determine the probability distributions $P(\psi_p|\pm)$ and $P(\bar{\psi}|\pm)$ in the presence of a stochastic turn-on time $t_i$. In order to extract a maximum of information associated with the qubit state, we need precise knowledge of these distributions. Indeed, the tails of the experimental distributions obtained for similar readouts~\cite{myerson2008,morello2010,dreau2013}, which determine the error rates \eqref{eq:conditionalErrorRates}, often strongly deviate from simple Gaussian-like behavior. These distributions can be found numerically from a Monte Carlo analysis of this model \cite{morello2010}. However, an analytical description is helpful in understanding the benefits of one post-processing scheme over another. Moreover, an analytical understanding of the statistics of the filters enables a fast extraction of the fidelity from the data, eliminating the need for time-consuming Monte-Carlo simulations. Therefore, in the following we derive exact analytical expressions for the peak-signal and boxcar distributions. Since the boxcar filter is a special case of the peak-signal filter, we first focus on obtaining $P(\psi_p|\pm)$.

\subsection{Probability distributions for a stochastic turn-on time}

\subsubsection{Peak-signal distribution \label{sec:peakDistribution}}

\begin{figure}
\centering
\includegraphics[width=\columnwidth]{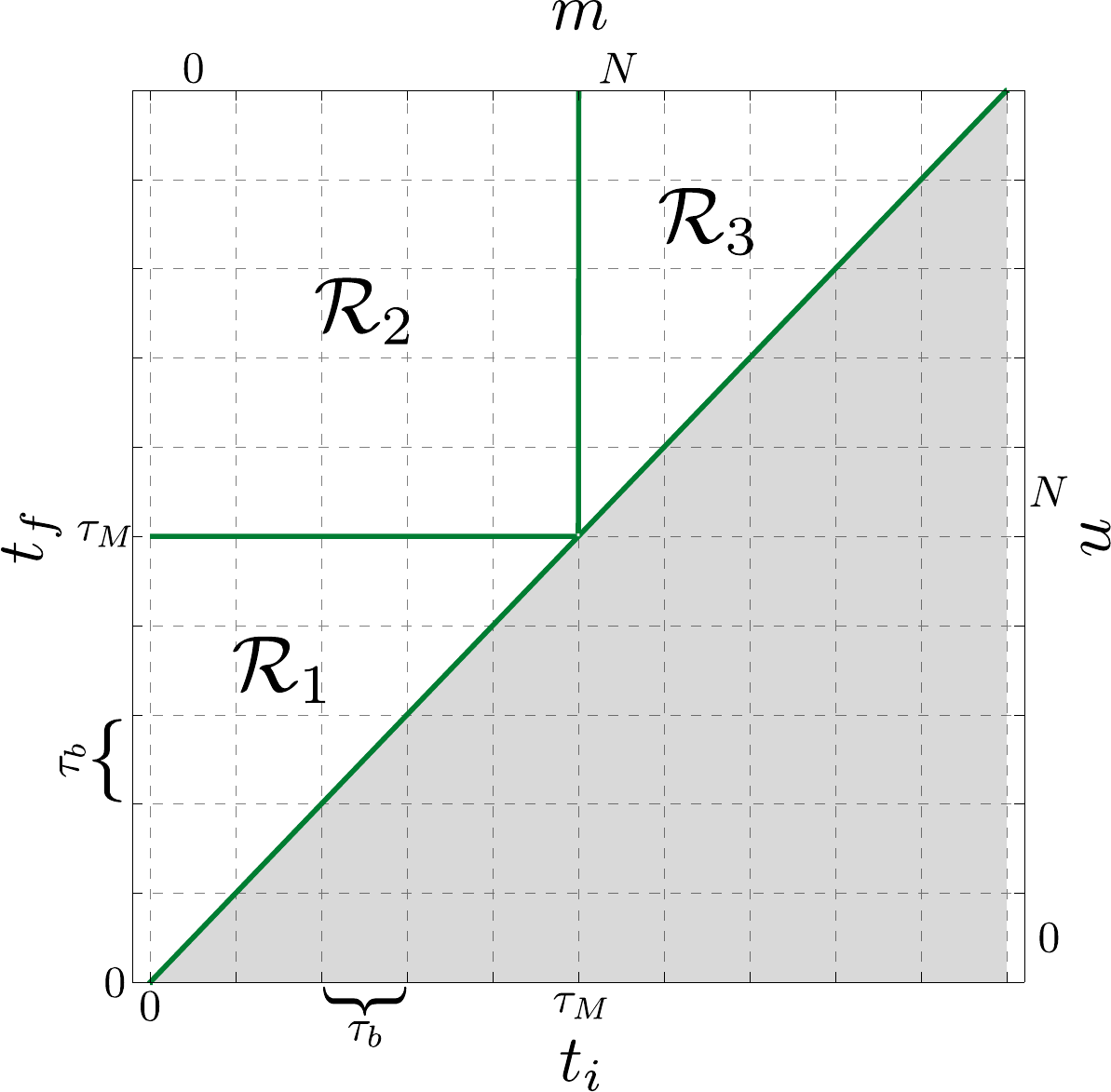}
\centering
\caption{(Color online) Regions of the $t_i- t_f$ plane. The turn-on time $t_i$ (turn-off time $t_f$) falls in the $m^{\mathrm{th}}$ ($n^{\mathrm{th}}$) discrete bin of length $\tau_b = \tau_M/N$, where $N$ is the number of bins contained in the measurement time $\tau_M$. Here, the $m^{\mathrm{th}}$ ($n^{\mathrm{th}}$) bin of the $t_i$ ($t_f$) axis starts at time $m \tau_b$ ($n \tau_b$). The shaded region is forbidden since we must necessarily have $t_f > t_i$. The finite measurement time $\tau_M$ divides the plane in three regions $\mathcal{R}_i$, Eq.~\eqref{eq:regions}. Each region gives a distinct, mutually-exclusive contribution to the peak-signal distribution $P(\psi_p|+)$.\label{fig:fig2}}
\end{figure}

As illustrated in Fig.~\ref{fig:fig2}, the turn-on time $t_i$ and the turn-off time $t_f$ must each fall in a random bin of length $\tau_b$~(see Sec.~\ref{sec:modelFilters}). In the following, we assume that $t_i$ ($t_f$) falls in the $m^{\mathrm{th}}$ ($n^{\mathrm{th}}$) bin. Therefore, using Bayes' rule to account for all possibilities, we write the peak-signal distributions as:
\begin{align}
	P(\psi_p|\pm)=\sum_{m=0}^{\infty}\sum_{n=m}^{\infty} P(\psi_p|\pm;m,n) P(m,n), \label{eq:bayesRulemn}
\end{align}
where $P(m,n)$ is the probability that $t_i$ and $t_f$ fall in bins $m$ and $n\ge m$, respectively:
\begin{align}
	P(m,n)=\int_m\!\!\!\!dt_i\int_n\!\! dt_f\, P(t_i,t_f). \label{eq:probabilitymn}
\end{align}
In the last expression, the integrals are taken over the square labeled by $(m,n)$ in the $t_i-t_f$ plane (see Fig.~\ref{fig:fig2}). Note that we allow for the possibility that $t_i$ and $t_f$ fall outside the measurement window ($m\ge N$ and $n\ge N$). Likewise, $P(\psi_p|\pm;m,n)$ is the probability distribution for $\psi_p$ conditional on $t_i$ and $t_f$ falling in a given square $(m,n)$:
\begin{align}
	P(\psi_p&|\pm;m,n) \notag \\
	&=\int_m\!\!\!\! dt_i\int_n\!\! dt_f\,P(\psi_p|\pm;t_i,t_f)P(t_i,t_f|m,n), \label{eq:peakConditionalmn}
\end{align}
i.e. it is the average of the distribution $P(\psi_p|\pm;t_i,t_f)$ over a square $(m,n)$ of the $t_i-t_f$ plane. Here, $P(t_i,t_f|m,n)$ is the distribution \eqref{eq:probabilitytitf} renormalized so that $t_i$ and $t_f$ lie in the cell $(m,n)$. We proceed to evaluate expressions \eqref{eq:probabilitymn} and \eqref{eq:peakConditionalmn}, which we then substitute into Eq.~\eqref{eq:bayesRulemn}.

First, we obtain $P(m,n)$, Eq.~\eqref{eq:probabilitymn}, by direct integration of Eq.~\eqref{eq:probabilitytitf}. Unsurprisingly, we find the discrete counterpart to the exponential form \eqref{eq:probabilitytitf}:
\begin{align}
	P(m,n)=
	\left\{
	\begin{array}{ll}
		\mathcal{D}_{if} e^{-\Gamma \tau_b m} & \mathrm{if}\;m=n , \\
		\mathcal{D}_i \mathcal{D}_f e^{-\Gamma \tau_b m} e^{-\tau_b(n-m)} & \mathrm{if}\;n> m ,
	\end{array}
	\right. \label{eq:probabilitymnExplicit}
\end{align}
where we define the normalization constants:
\begin{align}
	&\mathcal{D}_{if}=\frac{\Gamma(1-e^{-\tau_b})-(1-e^{-\Gamma \tau_b})}{\Gamma-1}, \notag \\
	&\mathcal{D}_i=\frac{\Gamma\left(1-e^{-(\Gamma-1)\tau_b}\right)}{\Gamma-1}, \label{eq:normalization} \\
	&\mathcal{D}_f=1-e^{-\tau_b}. \notag
\end{align}

Next, we derive the probability distributions $P(\psi_p|\pm;m,n)$, Eq.~\eqref{eq:peakConditionalmn}. Using the definition \eqref{eq:peakSignal} of the peak signal and a combinatorial argument, we show in Appendix~\ref{app:combinatorial} that the peak-signal distributions for fixed $t_i$ and $t_f$ are given by:
\begin{align}
	\begin{split}
	P(\psi_p&|\pm;t_i,t_f)= \\
	&\left(\prod_{S_{\phi}} q_{\phi}(\psi_p)^{N_{\phi}}\right)\times\left(\sum_{S_{\phi}} N_{\phi} \frac{p_{\phi}(\psi_p)}{q_{\phi}(\psi_p)}\right), \label{eq:combinatorialDistribution}
	\end{split}
\end{align}
where $S_{\phi}=\left\{l<N|\left\langle\bar{\psi}_l\right\rangle=\phi\right\}$ is the subset of $N_{\phi}$ bins in the measurement window $(0<t<\tau_\mathrm{M})$ having identical distributions $p_{\phi}$ and $q_{\phi}$, Eqs.~\eqref{eq:binDistribution} and \eqref{eq:cumulativeBinDistribution}, with average signal $\phi=\left\langle\bar{\psi}_l\right\rangle$. We note that $\sum_{S_{\phi}}N_{\phi} = N$.

To illustrate Eq.~\eqref{eq:combinatorialDistribution}, first assume that the qubit state is $\ket{-}$. In this case, all $N$ bins have the same average signal $\phi=-1$. Thus, Eq.~\eqref{eq:combinatorialDistribution} contains a single term:
\begin{align}
	P(\psi_p|-;t_i,t_f) = q_{-}^N \times N \frac{p_{-}}{q_{-}}=N q_{-}^{N-1} p_{-},
\end{align}
where $p_{-}(\psi_p)=\mathcal{N}_{\sigma}(\psi_p+1)$, Eq.~\eqref{eq:binDistribution}. Since the peak signal manifestly does not depend on $t_i$ and $t_f$ when the state is $\ket{-}$, the average \eqref{eq:peakConditionalmn} and the sum \eqref{eq:bayesRulemn} are trivial and we obtain:
\begin{align}
	P(\psi_p|-)=N q_{-}(\psi_p)^{N-1} p_{-}(\psi_p). \label{eq:peakDistributionMinus}
\end{align}

In the case where the initial qubit state is $\ket{+}$, the distribution \eqref{eq:combinatorialDistribution} takes a different form in each of the regions $\mathcal{R}_i$ of the $t_i-t_f$ plane depicted in Fig.~\ref{fig:fig2}:
\begin{align}
\begin{split}
\mathcal{R}_1:\; t_i < \tau_M \; , \; t_f < \tau_M; \\
\mathcal{R}_2:\; t_i < \tau_M \; , \; t_f > \tau_M; \label{eq:regions}\\
\mathcal{R}_3:\; t_i > \tau_M \; , \; t_f > \tau_M.
\end{split}
\end{align}
As an example, consider the case where $t_i$ and $t_f$ fall in region $\mathcal{R}_1$. If $t_i$ and $t_f$ fall in the same bin $m=n$, bin $m$ has an average signal $\phi=x\equiv 2(t_f-t_i)/\tau_b -1$ and the remaining $N_{-}=N-1$ bins have $\phi=-1$. Thus, in this particular case, Eq.~\eqref{eq:combinatorialDistribution} takes the form:
\begin{align}
	\begin{split}
	P(\psi_p|+;& t_i,t_f)= \\
	&q_{-}^{N-1} q_x \times\left[(N-1)\frac{p_{-}}{q_{-}}+\frac{p_x}{q_x}\right], \label{eq:peakRegion1Equal}
	\end{split}
\end{align}
where $p_x(\psi_p)=\mathcal{N}_{\sigma}(\psi_p-x)$, Eq.~\eqref{eq:binDistribution}. Substituting Eq.~\eqref{eq:peakRegion1Equal} into Eq.~\eqref{eq:peakConditionalmn}, we obtain:
\begin{align}
	\begin{split}
	P(\psi_p|+;& m,m)= \\
	&q_{-}^{N-1} \bar{q}_{if}\times\left[(N-1)\frac{p_{-}}{q_{-}}+\frac{\bar{p}_{if}}{\bar{q}_{if}}\right], \label{eq:peakRegion1Diff}
	\end{split}
\end{align}
where 
\begin{equation}
\bar{p}_{if}(\psi_p)=\int_m\!\! dt_i\int_{n=m}\!\! dt_f\;p_{x}(\psi_p)\cdot P(t_i,t_f|m,n)
\end{equation}
is the average probability distribution in a bin containing both $t_i$ and $t_f$. Similarly, when $t_i$ and $t_f$ fall in different bins ($m<n$), bin $m$ has $\phi=y\equiv 1-2(t_i-m\tau_b)/\tau_b$ and bin $n$ has $\phi=z\equiv 2(t_f-n \tau_b)/\tau_b -1$. Of the remaining bins, there are $N_{-}=N-(n-m)-1$ with $\phi=-1$ and $N_{+}=(n-m)-1$ with $\phi=+1$. Thus, we find:
\begin{align}
	\begin{split}
P(&\psi_p|+;m<n)=\\
&q_{-}^{N_{-}} q_{+}^{N_{+}}\bar{q}_i \bar{q}_f \times\left[N_{-}\frac{p_{-}}{q_{-}}+N_{+}\frac{p_{+}}{q_{+}}+\frac{\bar{p}_i}{\bar{q}_i}+\frac{\bar{p}_f}{\bar{q}_f}\right],
	\end{split}
\end{align}
where 
\begin{equation}
\bar{p}_{i(f)}(\psi_p)=\int_m\!\! dt_i\int_{n\neq m}\!\! dt_f\;p_{y(z)}(\psi_p)\cdot P(t_i,t_f|m,n)
\end{equation} 
is the average probability distribution in a bin containing only $t_i\,(t_f)$.

In Appendix~\ref{app:analyticalDistributionPlus}, we give similar expressions for $P(\psi_p|+;m,n)$ in regions $\mathcal{R}_2$ and $\mathcal{R}_3$ [Eqs.~\eqref{eq:peakConditionalmnExplicitPlus2} and \eqref{eq:peakConditionalmnExplicitPlus3}] as well as analytical expressions for the distributions $\bar{p}_{if}$, $\bar{p}_{i}$ and $\bar{p}_{f}$ and their respective cumulative distributions [Eqs.~\eqref{eq:barpif}, \eqref{eq:barpi} and \eqref{eq:barpf}]. We then perform the sum \eqref{eq:bayesRulemn} analytically and find that the probability distribution $P(\psi_p|+)$ has a contribution from each region of Fig.~\ref{fig:fig2}:
\begin{align}
	P(\psi_p|+)=P_1(\psi_p|+)+P_2(\psi_p|+)+P_3(\psi_p|+). \label{eq:peakDistributionPlus}
\end{align}
The contribution from region $\mathcal{R}_3$ arises from events in which the entire pulse occurs outside the measurement window, causing additional errors that could not occur if the turn-on time were deterministic ($t_i\rightarrow0$). Explicit expressions for each term in Eq.~\eqref{eq:peakDistributionPlus} are given in Eqs.~\eqref{eq:peakDistributionRegion1} and \eqref{eq:peakDistributionExplicit} of Appendix~\ref{app:analyticalDistributionPlus}.

In Fig.~\ref{fig:fig3}, we fit the analytical expressions \eqref{eq:peakDistributionMinus} and \eqref{eq:peakDistributionPlus} to the experimental data from the spin-to-charge conversion readout of Ref.~\cite{morello2010}. We find the fitted values of the SET current $I$ and of the prior probabilities $P(\pm)$ to be in good agreement with the measured values. The theoretical probability distributions provide a good fit to the data, allowing for a fast extraction of the readout error rate. Moreover, the importance of describing the probability distributions with precision is apparent from the non-Gaussian features of the distributions in Fig.~\ref{fig:fig3}. Indeed, both the protuberance in the left tail of $P(\psi_p|+)$, which is masked by $P(\psi_p|-)$ in an experiment, and the asymmetry in the distribution $P(\psi_p|+)$ must be accurately described to obtain a genuine estimate of the error rates \eqref{eq:conditionalErrorRates}.
\begin{figure}
\centering
\includegraphics[width=\columnwidth]{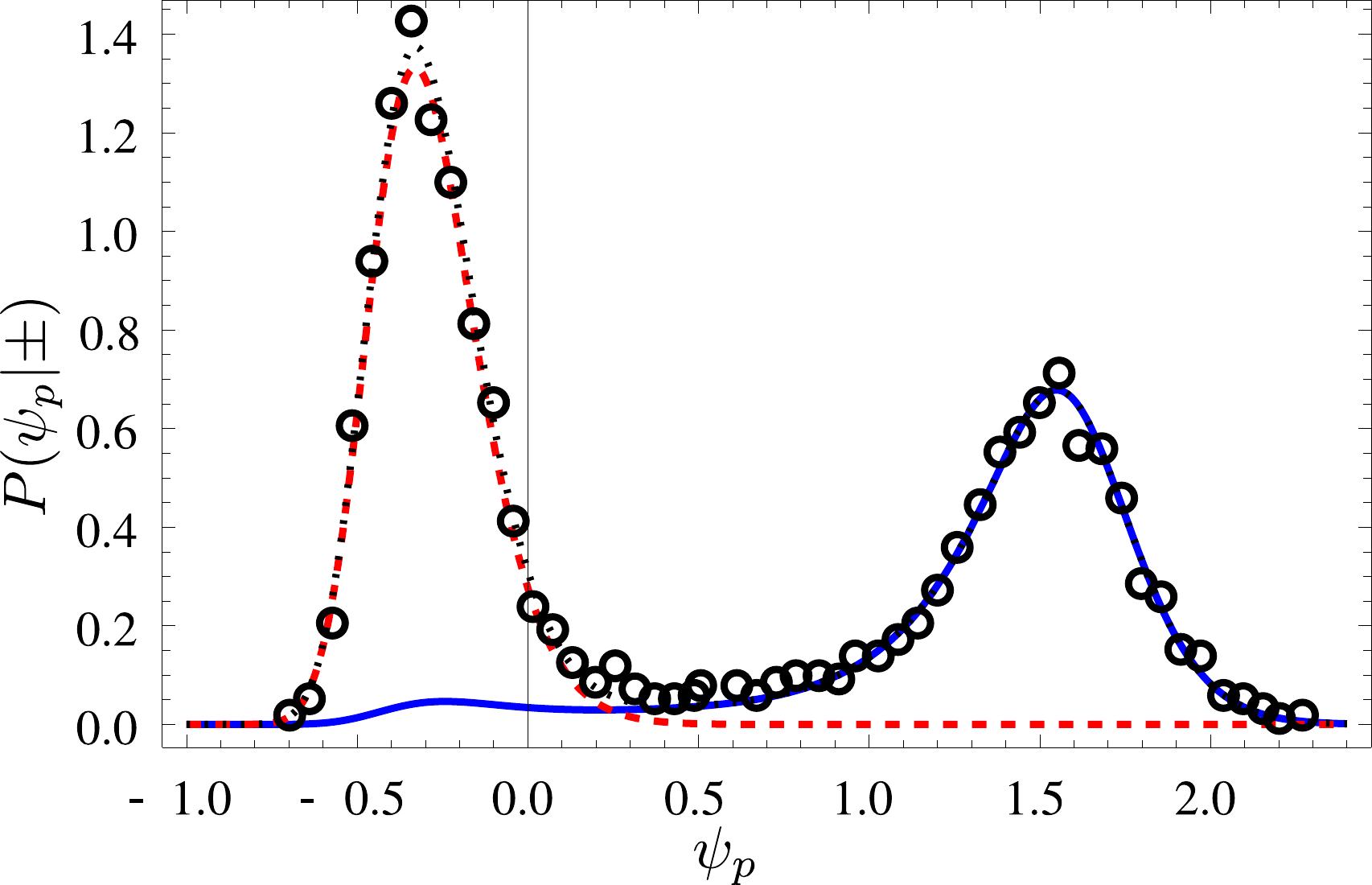}
\centering
\caption{(Color online) Fit of the analytical distributions $P(\psi_p|-)P(-)$ (dashed red line) and $P(\psi_p|+)P(+)$ (solid blue line), Eqs.~\eqref{eq:peakDistributionMinus} and \eqref{eq:peakDistributionPlus}, to the experimentally determined distribution $P(\psi_p)$ extracted from data in Fig.~4(b) of Ref.~\cite{morello2010} (open circles). The dotted black line is the full peak-signal distribution $P(\psi_p)=P(\psi_p|-)P(-)+P(\psi_p|+)P(+)$. The peak SET current $I_p$ for the spin-to-charge conversion is mapped to the reduced peak signal $\psi_p$ via $\psi_p=(2 I_p - I)/I$, where $I$ is the average SET current. We set $\Gamma$ and $\tau_M$ to their measured values $\Gamma=4$ and $\tau_M=2.5$ and we find fitted values of $I\approx2.0\,\mathrm{nA}$, $r\approx110$, $\tau_b\approx0.075$ and $P(+)=1-P(-)=0.47$ for the current, signal-to-noise ratio, bin time and prior probabilities respectively. The values of $I$ and $P(\pm)$ are in good agreement with the experimentally measured values of $I^{\mathrm{exp}}\approx 1.9\,\mathrm{nA}$ and $P(+)^{\mathrm{exp}}\approx 0.47$. Furthermore, the bin time is of the same order of magnitude as the inverse bandwith of the low-pass filter used in Ref.~\cite{morello2010}, corresponding to $\tau_b^{\mathrm{exp}}\approx 0.2$. If the data were to be post-processed using the square binning defined in Eq.~\eqref{eq:binSignal}, our model could be used to obtain more accurate estimates of the bin time and signal-to-noise ratio. \label{fig:fig3}}
\end{figure}

\subsubsection{Boxcar-filter distribution}

The boxcar-filter distributions $P(\bar{\psi}|\pm)$ are obtained from Eqs.~\eqref{eq:peakDistributionMinus}, \eqref{eq:peakDistributionPlus}, \eqref{eq:peakDistributionRegion1} and \eqref{eq:peakDistributionExplicit} by setting $N=1$ ($\tau_b = \tau_M$) and $\psi_p = \bar{\psi}$:
\begin{align}
	\begin{split}
	P(\bar{\psi}|-) &= p_{-}, \\
	P(\bar{\psi}|+) &= \mathcal{D}_{if} \bar{p}_{if} + \mathcal{D}_i e^{-\tau_M} \bar{p}_i + e^{-\Gamma \tau_M}p_{-}. \label{eq:boxCarDistribution}
	\end{split}
\end{align}
Here, $\bar{p}_{if}$ and $\bar{p}_i$ are given by Eqs.~\eqref{eq:barpif} and \eqref{eq:barpi} with $\tau_b=\tau_M$. The first term of $P(\bar{\psi}|+)$ is the contribution from the case where both $t_i$ and $t_f$ fall within the measurement window (region $\mathcal{R}_1$ of Fig.~\ref{fig:fig2}). The second term comes from the case where only $t_i$ falls within the measurement window (region $\mathcal{R}_2$ of Fig.~\ref{fig:fig2}). The last term is the contribution coming from the possibility of the pulse occurring outside the measurement window (region $\mathcal{R}_3$ of Fig.~\ref{fig:fig2}).

\subsection{Limit of a deterministic turn-on time ($\Gamma\rightarrow\infty$) \label{sec:limitInf}}

\subsubsection{Peak-signal distribution}

In this section we obtain analytical expressions for the distributions $P(\psi_p|\pm)$ when the turn-on time is deterministic, $\Gamma\rightarrow\infty$. This is typically the relevant limit for fluorescence-based readouts \cite{myerson2008,neumann2010,vamivakas2010,robledo2011}, where the cycling process starts almost immediately at the beginning of the readout phase ($t_i\rightarrow 0$).

If the state is $\ket{-}$, the distribution \eqref{eq:peakDistributionMinus} is independent of $\Gamma$ and the peak-signal distribution remains unchanged:
\begin{align}
	P(\psi_p|-) = N q_{-}^{N-1} p_{-}. \label{eq:peakDistributionMinusInf}
\end{align}
Next, we consider the case where the state is $\ket{+}$. When $\Gamma\gg 1$ and $\Gamma\tau_b \gg 1$, the normalization constants \eqref{eq:normalization} simplify to:
\begin{align}
	\mathcal{D}_{if}\approx\mathcal{D}_{f}\approx 1-e^{-\tau_b} \; , \;\; \mathcal{D}_{i}\approx 1. \label{eq:limitNormalization}
\end{align}
Moreover, when the condition $\Gamma r^{-1} \gg \max\left\{1,\sigma\right\}$ is satisfied, we asymptotically expand the error functions in Eq.~\eqref{eq:functionsH} and obtain:
\begin{align}
	h(\psi,1-\Gamma)\approx \frac{2e^{\Gamma\tau_b}}{\Gamma \tau_b} p_{+}. \label{eq:limitFunctionsH}
\end{align}
Physically, the condition $\Gamma r^{-1} \gg \max\left\{1,\sigma\right\}$ corresponds to the requirement that the signal fluctuations on the time interval $\left\langle t_i \right\rangle$ be larger than the signal itself, making it impossible to resolve the jump occurring at $t_i$. Note that when $\sigma=(r \tau_b)^{-1/2}>1$, the condition $\Gamma r^{-1} \gg \max\left\{1,\sigma\right\}$ is already implied by $\Gamma\tau_b\gg 1$. From these considerations we see that, for any finite $\Gamma$, the effects due to the stochasticity of $t_i$ become relevant at sufficiently large signal-to-noise ratio $r$. Substituting Eqs.~\eqref{eq:limitNormalization} and \eqref{eq:limitFunctionsH} into Eqs.~\eqref{eq:barpif}, \eqref{eq:barpi} and \eqref{eq:barpf} and using $\Gamma\gg 1$ and $\Gamma\tau_b \gg 1$ once again, we obtain:
\begin{align}
	\begin{split}
	\bar{p}_{if} \approx \bar{p}_f &\approx \frac{\tau_b}{2(1-e^{-\tau_b})} h(\psi_p,1)\; , \;\;\bar{p}_i \approx p_{+}, \\
	\bar{q}_{if} \approx \bar{q}_f &\approx \frac{1}{1-e^{-\tau_b}} H(\psi_p,1) \; , \;\; \bar{q}_i \approx q_{+}, \label{eq:binDistributionExplicitInf}
	\end{split}
\end{align}
where $h(\psi_p,1)$ and $H(\psi_p,1)$ are given by Eq.~\eqref{eq:functionsH}. With these simplifications, the expressions \eqref{eq:peakDistributionExplicit} for $P(\psi_p|+)$ become:
\begin{align}
	\begin{split}
	P&_1^{m=n}(\psi_p|+)\approx \\
	 &\;\;\;\;\;(1-e^{-\tau_b})\; q_{-}^{N-1}\bar{q}_f \left[(N-1)\frac{p_{-}}{q_{-}}+\frac{\bar{p}_f}{\bar{q}_f}\right], \\
	 P&_1^{m<n}(\psi_p|+)\approx (1-e^{-\tau_b})f_{1}\left(e^{-\tau_b}\right),\\
	 P&_2(\psi_p|+)\approx e^{-N \tau_b} \; N q_{+}^N\;\frac{p_{+}}{q_{+}},\\
	 P&_3(\psi_p|+)\approx 0, \label{eq:peakDistributionExplicitInf}
	 \end{split}
\end{align}
where $f_1$ is given by Eq.~\eqref{eq:functionsF}. We see that in this limit, $P(\psi_p|+)$ has no contribution from region $\mathcal{R}_3$ since the pulse cannot fall outside the measurement window. Expressions for $P(\psi_p|\pm)$ in the limit $\Gamma\to\infty$ [Eqs.~\eqref{eq:peakDistributionMinusInf} and \eqref{eq:peakDistributionExplicitInf}] are plotted in Fig.~\ref{fig:fig4}(d).

\subsubsection{Boxcar filter distribution}
The boxcar filter for deterministic turn-on time, $\Gamma\rightarrow\infty$, is obtained as before by setting $N=1$ ($\tau_b=\tau_M$) and $\psi_p = \bar{\psi}$ in Eq.~\eqref{eq:peakDistributionExplicitInf}. We thus obtain the result derived in Ref.~\cite{gambetta2007}:
\begin{align}
	\begin{split}
	P(\bar{\psi}|-) &= p_{-},\\
	P(\bar{\psi}|+) &\approx (1-e^{-\tau_M}) \bar{p}_{f} + e^{-\tau_M} p_{+} , \label{eq:boxCarDistributionInf}
	\end{split}
\end{align}
where $\bar{p}_{f}$, Eq.~\eqref{eq:barpf}, is given by its limiting expression \eqref{eq:binDistributionExplicitInf} evaluated at $\tau_b=\tau_M$. The first term of $P(\bar{\psi}|+)$ comes from the case where $t_f$ falls within the measurement window, $t_f<\tau_M$, with probability $1-e^{-\tau_M}$ (region $\mathcal{R}_1$ of Fig.~\ref{fig:fig2}). The second term comes from the case where $t_f$ falls outside the measurement window, $t_f>\tau_M$, with probability $e^{-\tau_M}$ (region $\mathcal{R}_2$ of Fig.~\ref{fig:fig2}). Note again that there is no contribution from region $\mathcal{R}_3$ of Fig.~\ref{fig:fig2} since $t_i$ never falls outside the measurement window.

The conditional probability distributions $P(\bar{\psi}|\pm)$ given by Eq.~\eqref{eq:boxCarDistributionInf} are shown in Fig.~\ref{fig:fig4}(d) alongside the corresponding distributions for the peak-signal filter. There is relatively little qualitative difference between the distributions for the boxcar and peak-signal filters when $\Gamma\to\infty$ [Fig.~\ref{fig:fig4}(d)]. In contrast, it is clear that these two post-processing strategies generate very different conditional distributions when $\Gamma$ is finite [see Fig.~\ref{fig:fig4}(b)].

\section{Error rate for the peak-signal and boxcar filters \label{sec:errorRateAnalysis}}

\begin{figure*}
\centering
\includegraphics[width=\textwidth]{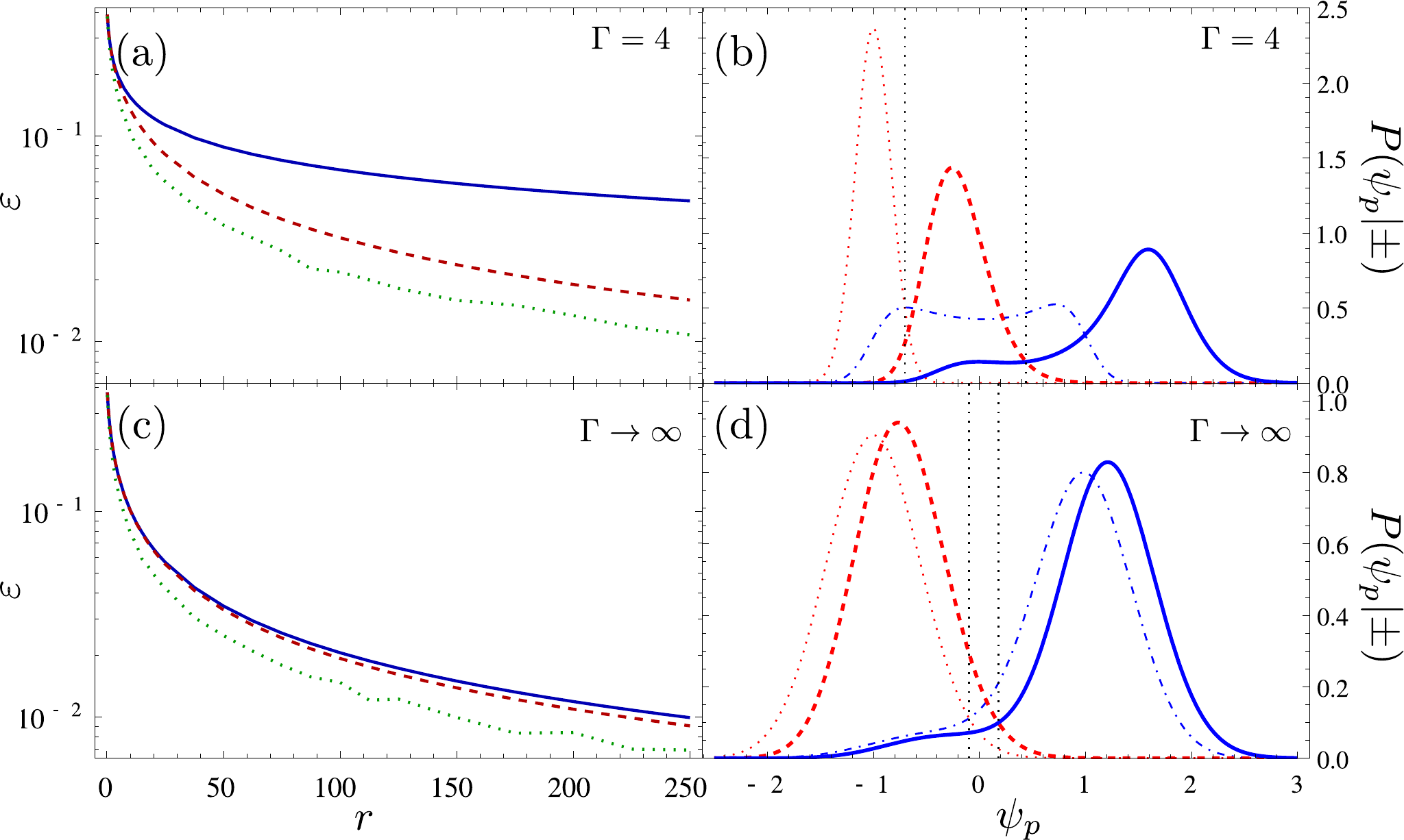}
\centering
\caption{(Color online) {\bf (a)},{\bf (c)} Optimized error rates, Eq.~\eqref{eq:averageErrorRate}, as a function of the signal-to-noise ratio $r$ for the boxcar filter (solid blue line), peak-signal filter (dashed red line) and maximum-likelihood filter (dotted green line) for {\bf (a)} $\Gamma=4$ (e.g. spin-to-charge conversion in semiconductor qubits) and {\bf (c)} $\Gamma\rightarrow\infty$ (e.g. fluorescence-based readouts in NV-centers or trapped ions). The error rates have been optimized with respect to the threshold $\nu$, measurement time $\tau_M$ and bin time $\tau_b$ (when applicable). In the case $\Gamma=4$, the error rate is significantly decreased for large $r$ by using the peak-signal filter instead of the boxcar filter, whereas the advantage is much smaller in the case $\Gamma\rightarrow\infty$. The error rate for the maximum-likelihood filter (dotted green line), obtained from the Monte-Carlo solution of Eqs.~\eqref{eq:stochasticEstimator} and \eqref{eq:stochasticEstimatorInf}, is the lowest theoretically achievable error rate. The fluctuations in the maximum-likelihood error rates are due to the finite sample size ($5\times 10^4$) of the Monte-Carlo simulation. {\bf (b)},{\bf (d)} Optimized probability distributions $P(\psi_p|-)$ (dashed red line), $P(\psi_p|+)$ (solid blue line), $P(\bar{\psi}|-)$ (dotted red line) and $P(\bar{\psi}|+)$ (dot-dashed blue line) for $r=30$ in the cases {\bf (b)} $\Gamma=4$, Eqs.~\eqref{eq:peakDistributionMinus}, \eqref{eq:peakDistributionPlus} and \eqref{eq:boxCarDistribution}, and {\bf (d)} $\Gamma\rightarrow\infty$, Eqs.~\eqref{eq:peakDistributionMinusInf}, \eqref{eq:peakDistributionExplicitInf} and \eqref{eq:boxCarDistributionInf}. For $\Gamma=4$, the weight of the distribution for $\ket{+}$ is visibly shifted to the right by using the peak-signal filter compared to the boxcar filter, decreasing the error rate significantly. When $\Gamma\rightarrow\infty$, the probability distributions for the peak-signal and boxcar filters are qualitatively the same in both cases and no advantage is gained. The dotted black vertical lines indicate the optimal threshold $\nu$ for each case, satisfying $P(\nu|-)=P(\nu|+)$. \label{fig:fig4}}
\end{figure*}
We are now in a position to compute the average error rate $\varepsilon$ for the peak-signal and boxcar filters. We first numerically integrate the analytical probability distributions derived in Sec.~\ref{sec:statistics} to obtain the conditional error rates \eqref{eq:conditionalErrorRates}. We then numerically minimize $\varepsilon$, Eq.~\eqref{eq:averageErrorRate}, with respect to the measurement time $\tau_M$, bin time $\tau_b$ and threshold $\nu$.

The optimized error rate is plotted as a function of the signal-to-noise ratio $r$ in Fig.~\ref{fig:fig4}(a) for the case of a stochastic turn-on time ($\Gamma=4$) \footnote{In addition to matching the experimental value found in Ref.~\cite{morello2010}, $\Gamma=4$ is a natural choice for a spin-to-charge conversion readout -- see the discussion in the final paragraph of Sec.~\ref{sec:bayesian}} and in Fig.~\ref{fig:fig4}(c) for the case of a deterministic turn-on time ($\Gamma\rightarrow\infty$). The advantage gained by measuring the peak signal instead of the time-averaged signal (employing the boxcar filter) is significant when $\Gamma=4$, but only marginal when $\Gamma\rightarrow\infty$. For example, when $\Gamma=4$, using the peak-signal instead of the boxcar filter increases the fidelity from $F=95.1\%$ to $F=98.4\%$ for $r=250$, whereas when $\Gamma\rightarrow\infty$, the fidelity only increases from $99.0\%$ to $99.1\%$ for the same signal-to-noise ratio. The qualitative difference between the two cases is apparent from the corresponding optimized probability distributions plotted in Figs.~\ref{fig:fig4}(b) and \ref{fig:fig4}(d) for a signal-to-noise ratio of $r=30$. In the case $\Gamma=4$, the weight of the probability distribution for $\ket{+}$ is shifted to higher values of $\psi_p$ by using the peak signal filter over the boxcar filter, whereas in the case $\Gamma\rightarrow\infty$, the peak-signal and boxcar distributions are qualitatively very similar.

We can better understand why this occurs by studying the asymptotic behavior of the error rate for the boxcar filter at large signal-to-noise ratio. Expression \eqref{eq:boxCarDistribution} for the boxcar filter probability distributions can be integrated analytically to give the unoptimized error rate:
\begin{align}
	\begin{split}
	\varepsilon = \frac{1}{2}\left[\mathcal{D}_{if}\bar{q}_{if}(\nu)\right. &+ \mathcal{D}_{i} e^{-\tau_M}\bar{q}_{i}(\nu)\\
	&\left.-(1-e^{-\Gamma\tau_M})q_{-}(\nu)+1\right]. \label{eq:boxCarAnalyticalErrorRate}
	\end{split}
\end{align}
The optimal threshold $\nu$ is given by the condition $P(\nu|+)=P(\nu|-)$. Thus, according to Eq.~\eqref{eq:boxCarDistribution}, $\nu$ is the solution of:
\begin{align}
	\mathcal{D}_{if}\bar{p}_{if}(\nu)+\mathcal{D}_{i}e^{-\tau_M}\bar{p}_{i}(\nu)=(1-e^{-\Gamma\tau_M})p_{-}(\nu). \label{eq:optimalThresholdCondition}
\end{align}
When the turn-on time $t_i$ is stochastic, there is a finite lower bound on the optimal measurement time $\tau_M$. Indeed, in such a case there is a finite probability that $t_i$ falls outside the measurement window. Therefore, we must necessarily choose an optimal measurement time $\tau_M \gtrsim 1$ to minimize the possibility of completely missing the pulse. This implies that as the signal-to-noise ratio increases, $r\rightarrow\infty$, the typical width of the distribution on the right-hand side of Eq.~\eqref{eq:optimalThresholdCondition} goes as $\sigma=(r \tau_M)^{-1/2}\rightarrow 0$ while the distribution on the left-hand side remains delocalized [see Eqs.~\eqref{eq:barpif} and \eqref{eq:barpf}]. Thus, the solution of Eq.~\eqref{eq:optimalThresholdCondition} must be such that the optimal threshold approaches $\nu \rightarrow -1$. Therefore, we expand the condition \eqref{eq:optimalThresholdCondition} asymptotically in the limit $\nu\rightarrow -1$ and $r\rightarrow\infty$ and find:
\begin{align}
	\nu\approx \sqrt{\frac{2}{r \tau_M}}\ln^{\frac{1}{2}}\left(\sqrt{\frac{2 r \tau_M}{\pi}}\gamma\right)-1, \label{eq:thresholdAsymptotic}
\end{align}
where:
\begin{align}
	\gamma=\frac{1}{\tau_M}\left(\frac{1-e^{-\Gamma\tau_M}}{1-(1-\Gamma)e^{-\Gamma\tau_M}}\right).
\end{align}
Next, we expand Eq.~\eqref{eq:boxCarAnalyticalErrorRate} in the same limit and use Eq.~\eqref{eq:thresholdAsymptotic} to obtain $\varepsilon$:
\begin{align}
	\begin{split}
	\varepsilon\approx \frac{1}{2}e^{-\Gamma\tau_M}&+\frac{1}{4}\sqrt{\frac{\tau_M}{r}}\left[1-(1-\Gamma)e^{-\Gamma\tau_M}\right] \\
	&\times \left[\ln^{\frac{1}{2}}\left(\frac{2\gamma^2 r \tau_M}{\pi}\right)+\ln^{-\frac{1}{2}}\left(4r\tau_M\right)\right]. \label{eq:errorRateExpansion}
	\end{split}
\end{align}
Since the first term decreases exponentially with $\tau_M$ and the second term increases polynomially with $\tau_M$, the optimal measurement time must diverge logarithmically when $r\rightarrow\infty$, $\tau_M \sim \ln r$. Thus, we use $e^{-\Gamma \tau_M}\ll 1$ and optimize Eq.~\eqref{eq:errorRateExpansion} with respect to $\tau_M$ when $r\rightarrow\infty$. We find the following leading logarithmic asymptotic form for the error rate:
\begin{align}
	\varepsilon \sim \frac{1}{\sqrt{r}}\ln r \quad (\Gamma<\infty).\label{eq:errorRateAsymptotic}
\end{align}
This result is to be compared to the case of a deterministic turn-on time $t_i$. In Ref.~\cite{gambetta2007}, it was shown that in this case, the average error rate for the boxcar filter scales instead as 
\begin{equation}
\varepsilon\sim \frac{1}{r}\ln r \quad (\Gamma\to\infty).
\end{equation} 
This qualitative difference in scaling arises from the fact that when $\Gamma\rightarrow\infty$, the optimal measurement time approaches $\tau_M \sim \varepsilon \sim \ln r / r \rightarrow 0$ when $r\rightarrow\infty$. This ensures that the turn-off time falls outside the measurement window, $t_f>\tau_M$, and thus that $\left\langle\bar{\psi}\right\rangle \approx +1$ when the state is $\ket{+}$ [see Fig.~\ref{fig:fig4}(d)]. As we already argued, this is not possible in the presence of a stochastic turn-on time since the optimal measurement time must always be such that $\tau_M\gtrsim 1$ in order to avoid the possibility of missing the pulse. Since the pulse can occur anywhere in the measurement window, we have $\left\langle\bar{\psi}\right\rangle<1$ when the state is $\ket{+}$, increasing the error rate [see Fig.~\ref{fig:fig4}(b)]. We have numerically verified that the two-time boxcar filter described in Sec.~\ref{sec:modelFilters} suffers from the same limitation. The peak-signal filter partly overcomes this shortcoming by gaining additional information on the location of the pulse within the measurement window, moving the average of the distribution back to $\left\langle\psi_p\right\rangle\gtrsim 1$ [see Fig.~\ref{fig:fig4}(b)].

\begin{figure}
\centering
\includegraphics[width=\columnwidth]{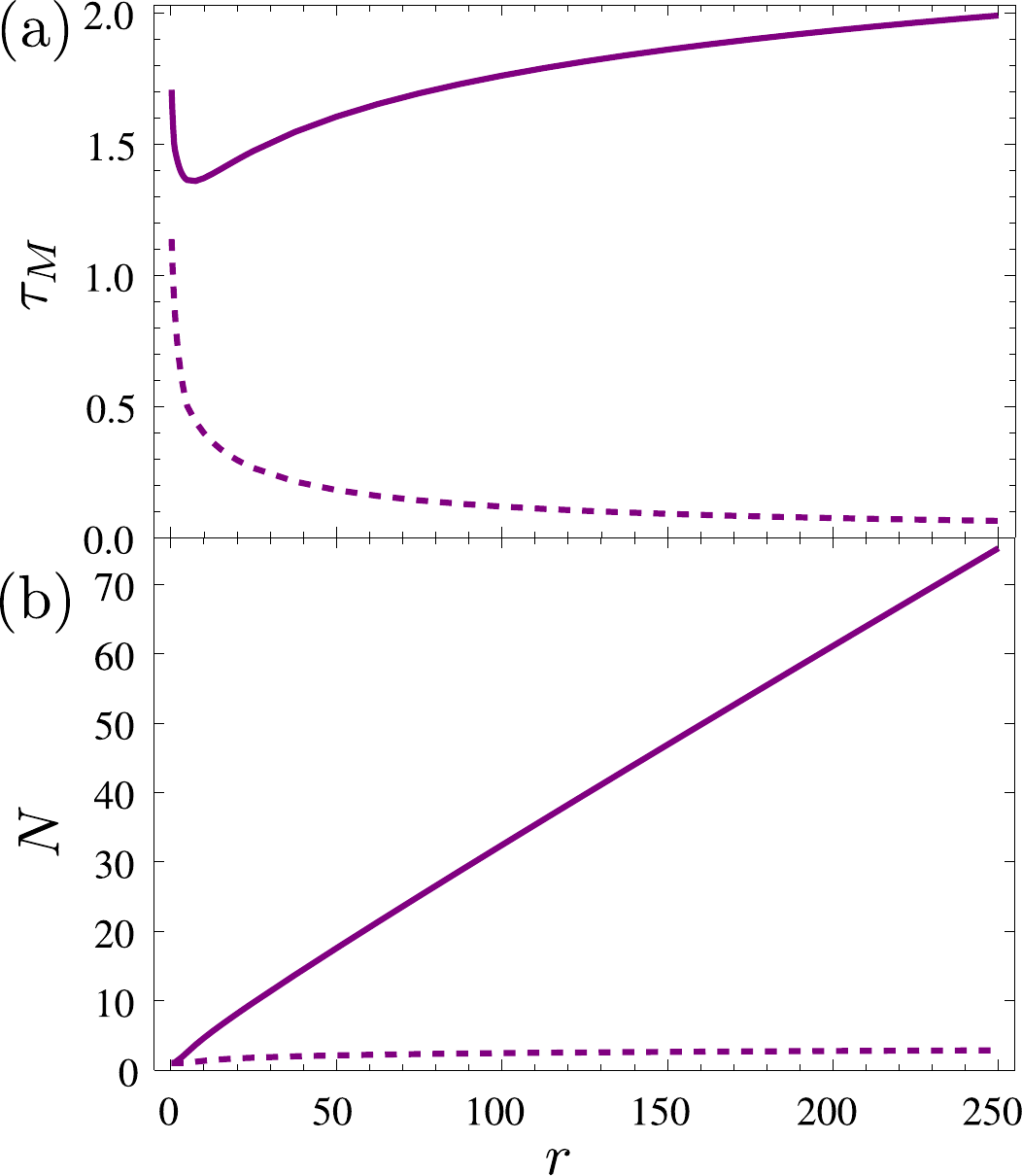}
\centering
\caption{(Color online) {\bf(a)} Optimal measurement time $\tau_M$ as a function of the signal-to-noise ratio $r$ for the peak-signal filter in the cases $\Gamma=4$ (purple solid line) and $\Gamma\rightarrow\infty$ (purple dashed line). When $\Gamma=4$, the measurement time diverges logarithmically with $r$ [see the discussion following Eq.~\eqref{eq:errorRateExpansion}], whereas when $\Gamma\rightarrow\infty$, $\tau_M$ approaches $0$ as $r$ increases. {\bf(b)} Optimal number number of bins $N=\tau_M/\tau_b$ as a function of the signal-to-noise ratio $r$ for the peak-signal filter in the cases $\Gamma=4$ (purple solid line) and $\Gamma\rightarrow\infty$ (purple dashed line). Although we have derived our model for $N\in\mathbb{N}$, we treat $N$ as a continuous variable for numerical optimization: fractional values of $N$ must be seen as an interpolation between integer values. Because $\tau_M$ must remain finite when $\Gamma=4$, it becomes advantageous to increase the number of bins in order to locate the pulse within the measurement window. When $\Gamma\rightarrow\infty$ ($t_i=0$), $\tau_M$ can approach $0$ as $r$ increases, eliminating the need for binning. \label{fig:fig5}}
\end{figure}

To better illustrate this effect, we plot the numerically optimized measurement time and number of bins $N=\tau_M/\tau_b$ for the peak-signal filter as a function of signal-to-noise ratio in Fig.~\ref{fig:fig5}, for both $\Gamma=4$ and $\Gamma\rightarrow\infty$. When $\Gamma=4$, it becomes advantageous to increase the number of bins as $r$ increases since the measurement time remains finite and the location of the pulse is unknown. When $\Gamma\rightarrow\infty$, the advantage gained by binning is not significant since the measurement time can become arbitrarily small. These results suggest an explanation for why the peak-signal filter is typically used for spin-to-charge conversion readouts using semiconductor qubits (having a stochastic turn-on time)~\cite{elzerman2004,morello2010,simmons2011}, whereas fluorescence-based readouts (having a deterministic turn-on time) typically rely on the simple boxcar filter~\cite{myerson2008,neumann2010,vamivakas2010,robledo2011}. More importantly, we emphasize that there is a crossover from $\Gamma\rightarrow\infty$ ($\Gamma r^{-1} \gg \max\left\{1,\sigma\right\}$) to $\Gamma < \infty$ ($\Gamma r^{-1} \ll \max\left\{1,\sigma\right\}$) as $r$ increases (see Sec.~\ref{sec:limitInf}). Thus, for any readout with finite $\Gamma$ it will become necessary to use the peak-signal filter instead of the boxcar filter as the signal-to-noise ratio improves. 

\section{Maximum likelihood filter \label{sec:bayesian}}

In the previous sections, we have shown that for readouts relying on a cycling process (e.g. spin-to-charge conversion in semiconductor qubits), the presence of a stochastic turn-on time for the cycling can significantly decrease the fidelity when simple filters are used. In this section, we generalize the maximum-likelihood filter developed in Ref.~\cite{gambetta2007} for a deterministic turn-on time to the case of a stochastic turn-on time. We show that even for this theoretically optimal Bayesian inference procedure, the fidelity of the readout can be significantly degraded by the uncertainty in the turn-on time.

The maximum-likelihood filter uses all the information contained in a given measurement record $\psi(t)$ to infer the state of the qubit. The likelihood ratio, Eq.~\eqref{eq:likelihoodRatio}, now takes the form:
\begin{align}
	\Lambda = \frac{P[\psi(t)|+]}{P[\psi(t)|-]}. \label{eq:likelihoodRatioBayesian}
\end{align}
When the qubit state is $\ket{-}$, the average signal is $\left\langle\psi(t)\right\rangle=-1$, Eq.~\eqref{eq:averageSignalMinus}, so that the probability distribution for $\psi(t)$ is:
\begin{align}
	P[\psi(t)|-]= A e^{-\int_{0}^{\tau_M}dt \frac{\left[\psi(t)+1\right]^2 r}{2}},
\end{align}
where $A$ is a normalization constant. When the qubit state is $\ket{+}$, the average signal for fixed $t_i$ and $t_f$, Eq.~\eqref{eq:averageSignalPlus}, is $\left\langle \psi(t) \right\rangle = 2\left[\theta\left(t-t_i\right)-\theta\left(t-t_f\right)\right]-1 \equiv i(t)$, so that the probability distribution for $\psi(t)$ is, using Bayes' rule:
\begin{align}
	\begin{split}
	P[\psi(t)&|+]=\\
	 A &\int_{0}^{\infty}\!\!\!\!dt_i \int_{t_i}^{\infty}\!\!\!\!dt_f P(t_i,t_f) e^{-\int_{0}^{\tau_M}dt \frac{\left[\psi(t)-i(t)\right]^2 r}{2}}.
	 \end{split}
\end{align}
Using these expressions, the likelihood ratio \eqref{eq:likelihoodRatioBayesian} can be rewritten as:
\begin{align}
	\Lambda = \int_{0}^{\infty}\!\!\!\!dt_i \int_{t_i}^{\infty}\!\!\!\!dt_f P(t_i,t_f) e^{\int_{0}^{\tau_M}dt\,\psi(t)\left[i(t)+1\right]r}.
\end{align}
The integral has a contribution from each domain illustrated in Fig.~\ref{fig:fig2}:
\begin{align}
	\Lambda = \Lambda_1 + \Lambda_2 + \Lambda_3, \label{eq:likelihoodRatioRegions}
\end{align}
where:
\begin{align}
	\begin{split}
	\Lambda_1 &= \int_{0}^{\tau_M}\!\!\!\!dt_i \int_{t_i}^{\tau_M}\!\!\!\!dt_f \; \Gamma e^{-(\Gamma-1) t_i}e^{-t_f} e^{\int_{t_i}^{t_f}dt\;2\psi(t)r}, \\
	\Lambda_2 &= e^{-\tau_M}\int_{0}^{\tau_M}\!\!\!\! dt_i \; \Gamma e^{-(\Gamma-1)t_i} e^{\int_{t_i}^{\tau_M}dt\;2\psi(t)r}, \\
	\Lambda_3 &= e^{-\Gamma \tau_M}. \label{eq:likelihoodRatioExplicit}
	\end{split}
\end{align}
In principle, we can evaluate these integrals numerically to obtain $\Lambda$ given a particular measurement record $\psi(t)$. If $\Lambda<1$, we declare the state to be $\ket{-}$ and if $\Lambda>1$, we declare the state to be $\ket{+}$. If $\Lambda=1$ we choose randomly by throwing an unbiased coin. However, we can avoid the triple integrals and the potentially large numerical values of $\Lambda$ in Eq.~\eqref{eq:likelihoodRatioExplicit} by using an equivalent set of stochastic differential equations~\cite{risken1984,gambetta2007} for the estimator $P[+|\psi(t)]=\Lambda/(1+\Lambda)$ (see Appendix \ref{app:stochasticEquations}). If $P[+|\psi(t)]<1/2$, we infer that the state is $\ket{-}$ and if $P[+|\psi(t)]>1/2$, we infer that the state is $\ket{+}$.

Although expressions for the case of deterministic turn-on time have already been given in Ref.~\cite{gambetta2007}, we reproduce them here in our notation for completeness. Taking the limit $\Gamma\rightarrow\infty$ in Eq.~\eqref{eq:likelihoodRatioExplicit}, we find that the likelihood ratio only has contributions from regions $\mathcal{R}_1$ and $\mathcal{R}_2$ in Fig.~\ref{fig:fig2}:
\begin{align}
	\Lambda = \Lambda_1 + \Lambda_2, \label{eq:likelihoodRatioRegionsInf}
\end{align}
where:
\begin{align}
	\Lambda_1 &= \int_{0}^{\tau_M}\!\!\!\! dt_f\; e^{-t_f} e^{\int_{0}^{t_f}dt\;2\psi(t) r},\\
	\Lambda_2 &= e^{-\tau_M} e^{\int_{0}^{\tau_M}dt\;2\psi(t) r}. \label{eq:likelihoodRatioExplicitInf}
\end{align}
A set of stochastic differential equations equivalent to these integrals is also given in Appendix \ref{app:stochasticEquations}.

\begin{figure}
\centering
\includegraphics[width=\columnwidth]{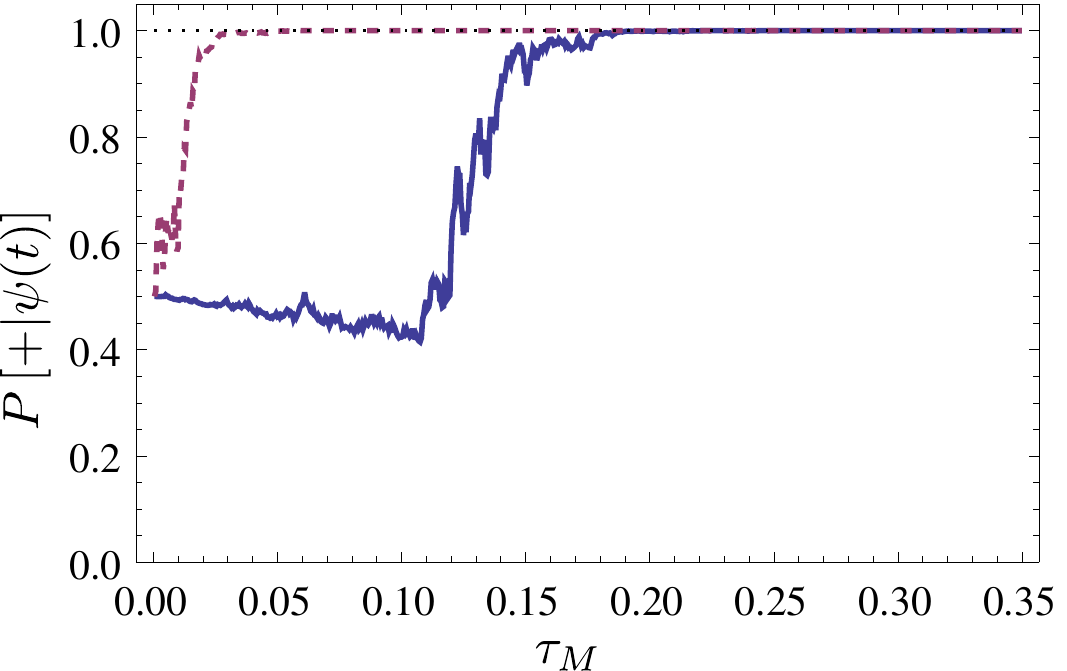}
\centering
\caption{(Color online) Estimator $P[+|\psi(t)]$ as a function of the measurement time $\tau_M$ for $r=30$, obtained for a randomly generated record $\psi(t)$ in the cases $\Gamma=4$ (solid blue line) and $\Gamma\rightarrow\infty$ (dashed purple line). When $\Gamma=4$, $P[+|\psi(t)]$ initially slowly decreases in the interval $[0,t_i]$ and suddenly jumps to $1$ when the pulse occurs. When $\Gamma\rightarrow\infty$, $t_i\rightarrow 0$ and $P[+|\psi(t)]$ immediatley jumps to $1$. In both cases the estimator reaches a constant value at large $\tau_M$ when all the available information on the pulse has been acquired. \label{fig:fig6}}
\end{figure}

To obtain the error rate of the maximum-likelihood filter, we generate $5\times10^4$ random records $\psi(t)$ by randomly choosing the initial state $\ket{\pm}$ with equal probability. For each record, we solve the stochastic differential equations \eqref{eq:stochasticEstimator} and \eqref{eq:stochasticEstimatorInf} using a standard fourth-order Runge-Kutta method~\cite{press2002} to obtain the estimator $P[+|\psi(t)]$. Typical solutions for $P[\psi(t)|+]$ as a function of $\tau_M$ are illustrated in Fig.~\ref{fig:fig6}. We see that the estimator reaches a constant value as $\tau_M$ increases. Thus, it is sufficient to choose a sufficiently large measurement time, $\tau_M\gg \left\langle t_f\right\rangle$ to obtain the optimal error rate~\cite{gambetta2007}. The average error rate $\varepsilon$ is then given by the fraction of records that are misidentified by the estimator. We plot $\varepsilon$ as a function of the signal-to-noise ratio $r$ in Fig.~\ref{fig:fig4}(a) for $\Gamma=4$ and in Fig.~\ref{fig:fig4}(b) for $\Gamma\rightarrow\infty$. The readout error rate is substantially larger at large $r$ when $\Gamma=4$ compared to $\Gamma\rightarrow\infty$. Quantitatively, we find that to achieve an error rate $\varepsilon < 1.1\%$ in the case $\Gamma=4$, it is necessary to have a signal-to-noise ratio of $r>250$. To achieve the same error rate when $\Gamma\to\infty$, it is sufficient to have a signal-to-noise ratio $r> 135$. Thus, we conclude that even for this optimal post-processing procedure, the additional uncertainty in $t_i$ can significantly degrade the single-shot fidelity of the readout.

An important consequence of this result is that it should always be possible to increase the fidelity of spin-to-charge conversion readouts by making the turn-on time deterministic in the sense of Sec.~\ref{sec:limitInf}. For example, suppose (similar to the experiment of Ref.~\cite{morello2010}) that an electron spin qubit in a localized orbital is coupled (with tunneling rate $\Gamma_0$) to each of $g$ nearly degenerate orbital states in a neighboring empty SET, initially in the Coulomb-blockade regime. The electron then tunnels from the excited spin state onto the SET at a rate $\Gamma_i= g\Gamma_0$, after which current can flow through the SET in the sequential tunneling regime, and the SET occupation $\rho_\mathrm{SET}(t)$ fluctuates between 1 and 0 electrons. An electron will tunnel back to the spin-qubit ground state at a $g$-independent rate $\Gamma_\tau\simeq \Gamma_0\bar{\rho}_\mathrm{SET}/2=\Gamma_0/4$ assuming the average SET occupation is $\bar{\rho}_\mathrm{SET}=1/2$ \footnote{$\bar{\rho}_\mathrm{SET}=1/2$ is realized when the tunneling rate to the drain lead ($\Gamma_d$) is balanced with the tunneling rate from the source lead ($\Gamma_s$), i.e. $\Gamma_s=\Gamma_d$. This choice has the added benefit of minimizing the contribution of shot noise to the signal-to-noise ratio.} and that the SET is occupied with spin-up and spin-down electrons with equal probability. The ratio of time scales setting $\Gamma$ is then $\left<\tau\right>/\left<t_i\right>=\Gamma=\Gamma_i/\Gamma_\tau= 4g$. The choice $\Gamma=4$ corresponds, in this case, to a single non-degenerate level of the unoccupied SET ($g=1$). Indeed, this happens to be the value found experimentally for the readout of Ref.~\cite{morello2010}. However, the degeneracy $g$ could be any value, in principle. The readout fidelity could be improved by increasing $g$ such that $\Gamma r^{-1}=4gr^{-1} > \max\left\{1,\sigma\right\}$, entering the regime where the asymptotic form, $\Gamma\to\infty$, of Fig. \ref{fig:fig4}(c) applies. In the typical case where the noise in an individual bin is small compared to the signal, $\sigma<1$, we find the very simple condition on the degeneracy $g$ and signal-to-noise ratio $r$:
\begin{equation}\label{eq:degeneracy-condition}
g > \frac{r}{4}.
\end{equation}
Nanostructures with a large density of single-particle states (e.g. a one-dimensional nanowire with a $E^{-1/2}$ singularity in the density of states), could be used to realize the limit given in Eq.~\eqref{eq:degeneracy-condition}, even in the limit of large signal-to-noise ratio $r$.

\section{Conclusions \label{sec:conclusions}}

In conclusion, we have shown that the fidelity of readouts relying on a QND cycling process with a stochastic turn-on time can be significantly increased by measuring the peak of the cycling signal (peak-signal filter) instead of its time average (boxcar filter). The origin of this discrepancy is that the peak-signal filter, by increasing the number of bins in the measurement window, can acquire additional information on the time at which the cycling signal occurs. Our results may explain why spin-to-charge conversion experiments in semiconductor qubits have typically used the peak-signal filter, whereas fluorescence-based readouts have normally relied on the simpler boxcar filter. Moreover, we predict that for any system with a stochastic turn-on time, however small, it will become advantageous to employ the peak-signal filter rather than the boxcar filter when the signal-to-noise ratio becomes larger than the dimensionless inverse average turn-on time ($r>\Gamma$). Furthermore, we have generalized the maximum-likelihood filter developed in Ref.~\cite{gambetta2007} to the case of a stochastic turn-on time. We have shown that even when this theoretically optimal procedure is followed, the presence of a stochastic turn-on time can significantly reduce the fidelity of the readout. Thus, we propose that the fidelity of such cycling readouts may be increased by making the turn-on time deterministic. In the case of a semiconductor qubit coupled to a nearby SET, this could be achieved by engineering the density of single-particle states for the SET to enhance tunneling from the qubit to the SET. It should be possible, in principle, to extend our approach to include shot-noise and non-Gaussian types of noise relevant in experiments with a small number of cycling events (e.g. in the presence of dark counts and near-Poissonian noise in ion-trap experiments~\cite{myerson2008}).

\section*{Acknowledgments}

The authors are indebted to J.P.~Dehollain and A.~Morello for sharing their data and experimental insight and to L.~Childress for useful discussions. We acknowledge financial support from NSERC, CIFAR, FRQNT, and INTRIQ.

\appendix

\section{Derivation of the general form of the peak-signal distribution $P(\psi_p|\pm;t_i,t_f)$\label{app:combinatorial}}

In this appendix, we derive the general form of the peak-signal distribution $P(\psi_p|\pm;t_i,t_f)$, Eq.~\eqref{eq:combinatorialDistribution}. For fixed turn-on time $t_i$ and turn-off time $t_f$, the probability of a given peak signal $\psi_p$ is the probability that at least one of the $N$ bins has $\psi_p<\bar{\psi}_l<\psi_p+d\psi_p$ while the remainder have $\bar{\psi}_l<\psi_p$:
\begin{align}
	\begin{split}
	P(&\psi_p|\pm;t_i,t_f)d\psi_p \\
	&=\prod_{S_{\phi}} \left[\sum_{k=0}^{N_{\phi}} B_{k}(p_{\phi},q_{\phi})\right] - \prod_{S_{\phi}} B_0(p_{\phi},q_{\phi}), \label{eq:combinatorialPeak}
	\end{split}
\end{align}
where $S_{\phi}=\left\{l<N|\left\langle\bar{\psi}_l\right\rangle=\phi\right\}$ runs over the subsets of $N_{\phi}$ bins in the measurement window having identical distributions $p_{\phi}$ and $q_{\phi}$, Eqs.~\eqref{eq:binDistribution} and \eqref{eq:cumulativeBinDistribution}. The total number of bins is $N = \sum_{S_{\phi}} N_{\phi}$. In Eq.~\eqref{eq:combinatorialPeak}, we introduced the binomial form:
\begin{align}
	\begin{split}
	B_{k}(p_{\phi}&,q_{\phi}) \equiv \binom{N_{\phi}}{k} \left[p_{\phi}(\psi_p) d\psi_p\right]^{k} \left[q_{\phi}(\psi_p)\right]^{N_{\phi}-k},
	\end{split}
\end{align}
which gives the probability that $k$ bins in the subset $S_{\phi}$ are such that $\psi_p<\bar{\psi}_l<\psi_p+d\psi_p$ and that $N_{\phi}-k$ bins are such that $\bar{\psi}_l<\psi_p$. Next, we perform the binomial sum in Eq.~\eqref{eq:combinatorialPeak} and find:
\begin{align}
	\begin{split}
	P(\psi_p|&\pm;t_i,t_f)d\psi_p = \\
	&\prod_{S_{\phi}}\left[p_{\phi}(\psi_p)d\psi_p+q_{\phi}(\psi_p)\right]^{N_{\phi}} - \prod_{S_{\phi}}q_{\phi}(\psi_p)^{N_{\phi}}. \label{eq:binomialSummed}
	\end{split}
\end{align}
In the continuum limit for $\psi_p$, we have $p_{\phi}(\psi_p)d\psi_p \ll q_{\phi}(\psi_p)$. Thus, we can expand the first term of Eq.~\eqref{eq:binomialSummed} to linear order to obtain the desired result, Eq.~\eqref{eq:combinatorialDistribution}:
\begin{align}
	\begin{split}
	P(\psi_p|\pm;&t_i,t_f)= \\
	 &\left(\prod_{S_{\phi}} q_{\phi}(\psi_p)^{N_{\phi}}\right)\times\left(\sum_{S_{\phi}} N_{\phi} \frac{p_{\phi}(\psi_p)}{q_{\phi}(\psi_p)}\right). \label{eq:combinatorialDistributionApp}
	\end{split}
\end{align}

\section{Analytical expressions for $P(\psi_p|+)$\label{app:analyticalDistributionPlus}}

In this appendix, we derive an explicit analytical expression for the distribution $P(\psi_p|+)$, Eq.~\eqref{eq:peakDistributionPlus}. Following the reasoning of Sec.~\ref{sec:peakDistribution}, we assume that the turn-on time $t_i$ and the turn-off time $t_f$ fall in the $m^{\mathrm{th}}$ and $n^{\mathrm{th}}$ time bin, respectively. We then write out Eq.~\eqref{eq:combinatorialDistributionApp} for $m$ and $n$ falling in each region of the $t_i-t_f$ plane depicted in Fig.~\ref{fig:fig2} and perform the average \eqref{eq:peakConditionalmn} over $t_i$ in bin $m$ and $t_f$ in bin $n$.

For $(m,n)$ in region $\mathcal{R}_1$, we find Eqs.~\eqref{eq:peakRegion1Equal} and \eqref{eq:peakRegion1Diff}:
\begin{align}
	\begin{split}
	P(\psi_p|&+;m,m)= \\
	&q_{-}^{N-1} \bar{q}_{if}\times\left[(N-1)\frac{p_{-}}{q_{-}}+\frac{\bar{p}_{if}}{\bar{q}_{if}}\right], \\
	P(\psi_p|&+;m<n)=q_{-}^{N-k-1} q_{+}^{k-1}\bar{q}_i \bar{q}_f \\
	\times&\left[(N-k-1)\frac{p_{-}}{q_{-}}+(k-1)\frac{p_{+}}{q_{+}}+\frac{\bar{p}_i}{\bar{q}_i}+\frac{\bar{p}_f}{\bar{q}_f}\right]. \label{eq:peakConditionalmnExplicitPlus1}
	\end{split}
\end{align}
where $k=n-m$. For $(m,n)$ in region $\mathcal{R}_2$, we find:
\begin{align}
	\begin{split}
	P(\psi_p|+;m,n)&= q_{-}^m q_{+}^{N-1-m} \bar{q}_i \\
	&\times \left[m \frac{p_{-}}{q_{-}}+(N-1-m)\frac{p_{+}}{q_{+}}+\frac{\bar{p}_i}{\bar{q}_i}\right]. \label{eq:peakConditionalmnExplicitPlus2}
	\end{split}
\end{align}
Finally, for $(m,n)$ in region $\mathcal{R}_3$, we find:
\begin{align}
	\begin{split}
	P&(\psi_p|+;m,n)= N q_{-}^{N-1} p_{-}, \label{eq:peakConditionalmnExplicitPlus3}
	\end{split}
\end{align}
Here, $\bar{p}_{if}$ is the average probability distribution in a bin that contains both $t_i$ and $t_f$ and $p_i\,(p_f)$ is the average probability distribution in a bin that contains only $t_i\,(t_f)$:
\begin{align}
	\begin{split}
	\bar{p}_{if}(\psi_p)&=\int_m\!\!\!\! dt_i\int_{n=m}\!\! dt_f\;p_{x}(\psi_p)\cdot P(t_i,t_f|m,n), \\
	\bar{p}_{i}(\psi_p)&=\int_m\!\!\!\! dt_i\int_{n\neq m}\!\! dt_f\;p_{y}(\psi_p)\cdot P(t_i,t_f|m,n), \\
	\bar{p}_{f}(\psi_p)&=\int_m\!\!\!\! dt_i\int_{n\neq m}\!\! dt_f\;p_{z}(\psi_p) \cdot P(t_i,t_f|m,n). \label{eq:binDistributionAveragedApp}
	\end{split}
\end{align}
Above, the integrals are taken over the square $(m,n)$ in the $t_i-t_f$ plane, Fig.~\ref{fig:fig2}. In Eq.~\eqref{eq:binDistributionAveragedApp}, $p_x$, $p_y$ and $p_z$ are the Gaussian distributions \eqref{eq:binDistribution} with average signal $x=2(t_f-t_i)/\tau_b -1$, $y=1-2(t_i-m\tau_b)/\tau_b$ and $z=2(t_f-n\tau_b)/\tau_b - 1$, respectively.

Using Eqs.~\eqref{eq:probabilitytitf} and \eqref{eq:binDistribution} to perform the integrals \eqref{eq:binDistributionAveragedApp}, we find that, in a bin that contains both $t_i$ and $t_f$ ($m=n$), the average probability distribution and its cumulative function are:
\begin{align}
	\begin{split}
	\bar{p}_{if}&=\frac{\tau_b}{2\mathcal{D}_{if}}\left[h(\psi_p,1)-e^{-\Gamma\tau_b} h(\psi_p,1-\Gamma)\right], \\
	\bar{q}_{if}&=\frac{1}{\mathcal{D}_{if}}\left[H(\psi_p,1)-\frac{e^{-\Gamma \tau_b}}{1-\Gamma}H(\psi_p,1-\Gamma)\right]. \label{eq:barpif}
	\end{split}
\end{align}
In a bin that contains only the turn-on time $t_i$, we find the average distributions:
\begin{align}
	\begin{split}
	\bar{p}_i &= \frac{\Gamma \tau_b}{2\mathcal{D}_i} e^{-(\Gamma-1)\tau_b}h(\psi_p,1-\Gamma), \\
	\bar{q}_i &= \frac{\Gamma}{(1-\Gamma)\mathcal{D}_i} e^{-(\Gamma-1)\tau_b}H(\psi_p,1-\Gamma), \label{eq:barpi}
	\end{split}
\end{align}
and in a bin that contains only the turn-off time $t_f$, they are:
\begin{align}
	\begin{split}
	\bar{p}_f &= \frac{\tau_b}{2\mathcal{D}_f} h(\psi_p,1), \\
	\bar{q}_f &= \frac{1}{\mathcal{D}_f} H(\psi_p,1). \label{eq:barpf}
	\end{split}
\end{align}
Here, we introduce the functions:
\begin{align}
	\begin{split}
	h(\psi,&\alpha)=\frac{1}{2}e^{\frac{\alpha^2 \tau_b}{8 r}-\frac{\alpha\tau_b(\psi+1)}{2}}\times \\
	&\left[\mathrm{erf}\left(\frac{\psi+1-\frac{\alpha}{2r}}{\sqrt{2\sigma^2}}\right)-\mathrm{erf}\left(\frac{\psi-1-\frac{\alpha}{2r}}{\sqrt{2\sigma^2}}\right)\right], \\
	H(\psi,&\alpha)=q_{-}-e^{-\alpha \tau_b}q_{+}-h(\psi,\alpha), \label{eq:functionsH}
	\end{split}
\end{align}
where $p_{\pm}(\psi_p)=\mathcal{N}_{\sigma}(\psi_p \mp 1)$ and $\sigma^2 = (r \tau_b)^{-1}$, Eq.~\eqref{eq:binDistribution}.

Next, we substitute Eqs.~\eqref{eq:probabilitymnExplicit}, \eqref{eq:peakConditionalmnExplicitPlus1}, \eqref{eq:peakConditionalmnExplicitPlus2} and \eqref{eq:peakConditionalmnExplicitPlus3} into Eq.~\eqref{eq:bayesRulemn} to obtain $P(\psi_p|+)$. We find that $P(\psi_p|+)$ has a contribution from each region $\mathcal{R}_i$:
\begin{align}
	P(\psi_p|+)=P_1(\psi_p|+)+P_2(\psi_p|+)+P_3(\psi_p|+), \label{eq:peakDistributionPlusApp}
\end{align}
where the contribution from region $\mathcal{R}_1$ has distinct contributions from $m=n$ and $m<n$:
\begin{align}
	P_1(\psi_p|+)=P_1^{m=n}(\psi_p|+)+P_1^{m<n}(\psi_p|+). \label{eq:peakDistributionRegion1}
\end{align}
We perform the sum \eqref{eq:bayesRulemn} directly and obtain an analytical form for $P(\psi_p|+)$:
\begin{align}
	P&_1^{m=n}(\psi_p|+)= \notag \\
	 &\mathcal{D}_{if}\;g_N\left(e^{-\Gamma\tau_b}\right)\; q_{-}^{N-1} \bar{q}_{if}\left[(N-1)\frac{p_{-}}{q_{-}}+\frac{\bar{p}_{if}}{\bar{q}_{if}}\right], \notag \\
	P&_1^{m<n}(\psi_p|+)= \label{eq:peakDistributionExplicit}\\
	 &\frac{\mathcal{D}_i\mathcal{D}_f}{1-e^{-\Gamma\tau_b}} \left[f_{1}\left(e^{-\tau_b}\right)-e^{-\Gamma \tau_b N}f_{1}\left(e^{-(1-\Gamma)\tau_b}\right)\right], \notag \\
	P&_2(\psi_p|+)=\frac{\mathcal{D}_i \mathcal{D}_f e^{-\tau_b N}}{1-e^{-\tau_b}}\;f_{2}\left(e^{-(\Gamma-1)\tau_b}\right), \notag \\
	P&_3(\psi_p|+)=e^{-\Gamma \tau_b N}\; q_{-}^N\; N \frac{p_{-}}{q_{-}}. \notag
\end{align}
Eq.~\eqref{eq:peakDistributionPlusApp}, together with Eqs.~\eqref{eq:peakDistributionRegion1} and \eqref{eq:peakDistributionExplicit}, is the central result of this appendix. Here, we have introduced the functions:
\begin{align}
	\begin{split}
	f_{1}(a)&=a\bar{q}_i \bar{q}_f \left[b_{-}\frac{p_{-}}{q_{-}}+b_{+}\frac{p_{+}}{q_{+}}+b_{if}\left(\frac{\bar{p}_{i}}{\bar{q}_{i}}+\frac{\bar{p}_{f}}{\bar{q}_{f}}\right)\right], \\
	f_{2}(a)&=\bar{q}_i\left(c_{-}\frac{p_{-}}{q_{-}}+c_{+}\frac{p_{+}}{q_{+}}+c_{i}\frac{\bar{p}_{i}}{\bar{q}_{i}}\right), \label{eq:functionsF}
	\end{split}
\end{align}
where the coefficients are given by:
\begin{align}
	\begin{split}
	b_{-}&= (q_{+} a)^{N-2} g'_{N-1}\left(\frac{q_{-}}{q_{+}a}\right), \\
	b_{+}&= q_{-}^{N-2} g'_{N-1}\left(\frac{q_{+}a}{q_{-}}\right), \\
	b_{if}&=q_{-}^{N-2} g_{N-1}\left(\frac{q_{+}a}{q_{-}}\right),
	\end{split}
\end{align}
and:
\begin{align}
	\begin{split}
	c_{-}&=q_{+}^{N-1} g'_N\left(\frac{q_{-} a}{q_{+}}\right), \\
	c_{+}&=(q_{-}a)^{N-1} g'_N\left(\frac{q_{+}}{q_{-}a}\right), \\
	c_{i}&=q_{+}^{N-1} g_N\left(\frac{q_{-}a}{q_{+}}\right).
	\end{split}
\end{align}
The functions $g$ and $g'$ are geometric sums arising from performing the sum \eqref{eq:bayesRulemn}:
\begin{align}
	\begin{split}
	g_N(u) &= \sum_{k=0}^{N-1} u^k = \frac{1-u^N}{1-u}, \\
	g'_N(u) &= \sum_{k=0}^{N-1} k u^k = \frac{u(1-u^N)-N u^N (1-u)}{(1-u)^2}.
	\end{split}
\end{align}

\section{Stochastic differential equations for the maximum-likelihood filter \label{app:stochasticEquations}}

In this appendix, we derive stochastic differential equations for the estimator $P[+|\psi(t)]$ plotted in Fig.~\ref{fig:fig6}. We first derive equations for the likelihood ratio $\Lambda$ and then reexpress them in terms of $P[+|\psi(t)]$.

We directly differentiate each member of Eq.~\eqref{eq:likelihoodRatioExplicit} with respect to $\tau_M$ to obtain a set of linear differential equations for $\Lambda=\Lambda_1+\Lambda_2+\Lambda_3$:
\begin{align}
	\begin{split}
	\frac{d\Lambda_1}{d\tau_M}&=\Lambda_2, \\
	\frac{d\Lambda_2}{d\tau_M}&=\Gamma\Lambda_3 + \left[2\psi(\tau_M) r - 1\right] \Lambda_2, \\
	\frac{d\Lambda_3}{d\tau_M}&=-\Gamma\Lambda_3. \label{eq:stochasticLambda}
	\end{split}
\end{align}
From Eq.~\eqref{eq:likelihoodRatioExplicit}, we see that these equations must be solved subject to the initial conditions $\Lambda_1(0)=0$, $\Lambda_2(0)=0$ and $\Lambda_3(0)=1$.

Using Eq.~\eqref{eq:likelihoodRatio}, we may express the estimator in terms of the likelihood ratio \eqref{eq:likelihoodRatioRegions} as:
\begin{align}
	P[+|\psi(t)]=\frac{\Lambda}{1+\Lambda}.
\end{align}
Defining $\mathpzc{P}_i=\Lambda_i/(1+\Lambda)$, we write the estimator as:
\begin{align}
	P[+|\psi(t)]=\mathpzc{P}_1 + \mathpzc{P}_2 + \mathpzc{P}_3.
\end{align}
In terms of the $\mathpzc{P}_i$'s, the equations \eqref{eq:stochasticLambda} transform into a set of non-linear, first-order differential equations:
\begin{align}
	\begin{split}
	\frac{d\mathpzc{P}_1}{d\tau_M}&=\mathpzc{P}_2 - 2\psi(\tau_M)r\,\mathpzc{P}_1 \mathpzc{P}_2, \\
	\frac{d\mathpzc{P}_2}{d\tau_M}&=\Gamma \mathpzc{P}_3 + \left[2\psi(\tau_M)r - 1\right]\mathpzc{P}_2 - 2\psi(\tau_M)r\,\mathpzc{P}_2^2, \\
	\frac{d\mathpzc{P}_3}{d\tau_M}&=-\Gamma\mathpzc{P}_3 - 2\psi(\tau_M)r \, \mathpzc{P}_2 \mathpzc{P}_3. \label{eq:stochasticEstimator}
	\end{split}
\end{align}
These equations must be solve with the initial conditions $\mathpzc{P}_1(0)=0$, $\mathpzc{P}_2(0)=0$ and $\mathpzc{P}_3(0)=1/2$. Note that when $r=0$, $dP[+|\psi(t)]/d\tau_M = 0$ for all $\tau_M$, and no information can be acquired on the qubit state.

When $\Gamma\rightarrow\infty$, we follow a similar procedure and find that the likelihood ratio $\Lambda=\Lambda_1+\Lambda_2$ is the solution of the following pair of equations~\cite{gambetta2007}:
\begin{align}
	\begin{split}
	\frac{d\Lambda_1}{d\tau_M} &= \Lambda_2, \\
	\frac{d\Lambda_2}{d\tau_M} &= \left[2\psi(\tau_M)r -1\right]\Lambda_2, \label{eq:stochasticLambdaInf}
	\end{split}
\end{align}
with the initial conditions $\Lambda_1(0)=0$ and $\Lambda_2(0)=1$. Similarly, the estimator $P[+|\psi(t)] = \mathpzc{P}_1 + \mathpzc{P}_2$ is the solution of:
\begin{align}
	\begin{split}
	\frac{d\mathpzc{P}_1}{d\tau_M} &= \mathpzc{P}_2 - 2\psi(\tau_M)r\,\mathpzc{P}_1 \mathpzc{P}_2, \\
	\frac{d\mathpzc{P}_2}{d\tau_M}&= \left[2\psi(\tau_M)r - 1\right]\mathpzc{P}_2 - 2\psi(\tau_M)r \,\mathpzc{P}_2^2, \label{eq:stochasticEstimatorInf}
	\end{split}
\end{align}
with the initial conditions $\mathpzc{P}_1(0)=0$ and $\mathpzc{P}_2(0)=1/2$.

\bibliography{pra.01}

\end{document}